\documentclass[a4paper,fleqn,usenatbib]{mnras}

\usepackage{newtxtext,newtxmath}

\usepackage[T1]{fontenc}
\usepackage{ae,aecompl}

\usepackage{graphicx}	
\usepackage{amsmath}	

\newcommand{\Eexc}{$E_{\rm exc}$}
\newcommand{\Teff}{T_{\rm eff}}
\newcommand{\logg}{\rm log~ g}
\newcommand{\kms}{km\,s$^{-1}$}
\newcommand{\eps}[1]{\log\varepsilon_{\rm #1}}
\newcommand{\eu}[5]{\mbox{$#1\,^#2{\rm #3}^{#4}_{\rm #5}$}}
\def\ione{\,{\sc i}}
\def\ii{\,{\sc ii}}
\def\iii{\,{\sc iii}}
\def\iv{\,{\sc iv}}

\title[Non-LTE spectroscopy of A-B stars]{Chemical diversity among A-B stars with low rotational velocities: non-LTE abundance analysis}

\author[L. Mashonkina et al.]{
L.~Mashonkina,$^{1}$\thanks{E-mail: lima@inasan.ru} T.~Ryabchikova,$^1$ S.~Alexeeva,$^{2}$ T.~Sitnova,$^{1}$ $\&$ O.~Zatsarinny$^{3}$ \\
$^{1}$Institute of Astronomy of the Russian Academy of Sciences, Pyatnitskaya st. 48, 119017, Moscow, Russia \\
$^{2}$CAS Key Laboratory of Optical Astronomy, National Astronomical Observatories, Beijing, 100101, China \\
$^{3}$Department of Physics and Astronomy, Drake University, Des Moines, Iowa 50311, USA
}

\date{Accepted XXX. Received YYY; in original form ZZZ}
\pubyear{2020}

\begin{document}
\label{firstpage}
\pagerange{\pageref{firstpage}--\pageref{lastpage}}
\maketitle

\begin{abstract}
We present accurate element abundance patterns based on the non-local thermodynamic equilibrium (non-LTE, NLTE) line formation for 14 chemical elements from He to Nd for a sample of nine A9 to B3 type stars with well determined atmospheric parameters and low rotational velocities. We constructed new model atom of Zr\ii -\iii\ and updated model atoms for Sr\ii\ and Ba\ii\ by implementing the photoionization cross sections from calculations with the Dirac $B$-spline $R$-matrix method. The NLTE abundances of He to Fe in the stars HD~17081, HD~32115, HD~160762, and HD~209459 are found to be consistent with the solar abundances, and HD~73666 being a Blue Straggler does not reveal deviations from chemical composition of the Praesepe cluster. Three of these stars with an effective temperature of lower than 10500~K have supersolar abundances of Sr, Zr, Ba, and Nd, and our results suggest the presence of a positive correlation between stellar effective temperature and abundance. For each star, enhancement of Ba is higher than that for any other heavy element. We propose that the solar Ba abundance is not representative of the galactic Ba abundance at modern epoch. 
The status of HD~145788 was not clarified: this star has solar abundances of C to Si and enhancements of Sr to Ba similar to that for superficially normal stars of similar temperature, while overabundant Ca, Ti, and Fe. 
The NLTE abundances of Vega support its status of a mild $\lambda$~Bootis star.
 
\end{abstract}




\section{Introduction}

The range of A to middle B spectral types is known by chemical anomalies of various kind. There exist magnetic (Ap), non-magnetic metallic-line (Am), non-magnetic mercury-manganese (HgMn), helium-weak (He-weak), $\lambda$~Bootis type stars. Most probably, only the surface layers of chemically peculiar (CP) stars possess chemical anomalies and their origin is connected with stellar magnetic fields and/or atomic diffusion that results from the competitive action of gravitational settling and radiative accelerations \citep{1970ApJ...160..641M,1980AJ.....85..589M}. For better understanding the mechanisms of chemical peculiarity, accurate abundance determinations are required for many elements, from the lighest He to the rare-earth elements (REE), in each of the group of CP stars. 

With all the uncertainties in theoretical spectrum modelling and analysis of observed spectra, the most reliable abundance results can be obtained for non-magnetic stars with low rotational velocities. As observations suggest \citep[][and references therein]{1974ARA&A..12..257P} and the theory supports \citep{1982ApJ...258..349M}, slow rotation (the equatorial velocity below 120~\kms) favours the presence of chemical anomalies in hot (A-B type) stars. Nevertheless, the A-B stars with close-to-solar element abundance pattern, which are referred to as superficially normal stars, exist, as shown by S.~Adelman and his collaborators in their series of papers \citep[][and references therein]{2000MNRAS.316..514A} and by the other researchers \citep{1990A&A...240..331L,1993AA...276..142H,1999A&A...351..247V,2002A&A...389..537B,2009AA...503..945F,2014A&A...562A..84R}. In order to make definite conclusions about abundance differences between superficially normal stars and the Sun and between superficially normal and Am stars, appropriate treatment of spectral line formation is required. However, the majority of abundance studies of A to late B type stars are still made under the LTE assumption. Exceptions are related to either selected stars or selected chemical elements. For example, for the benchmark A0~V star Vega, its abundances were derived based on the non-local thermal equilibrium (non-LTE = NLTE) line formation by \citet[][Fe]{1986A&A...165..170G}, \citet[][Mg, Ba]{Gigas88}, \citet[][C]{1991A&A...246..644S}, \citet[][Al]{1992LNP...401...57S}, \citet[][C, N]{1992PASJ...44..649T}, \citet[][O, Mg, C]{2000A&A...359.1085P,Przybilla_mg,2001A&A...379..936P}, and \citet[][N]{2001A&A...379..955P}. N.~Przybilla and his collaborators calculate the NLTE abundance patterns (He, C, N, O, Mg, S, Ti, Fe) for the selected A to late B supergiants \citep{2006A&A...445.1099P,2008A&A...479..849S}, the CP star HVS~7 \citep{2008A&A...488L..51P}, and the Am star HD~131399A \citep{2017A&A...604L...9P}.
The NLTE abundances of C, N, O in samples of A-type supergiants and bright giants were derived by \citet[][C and N]{1995ApJ...449..839V}, \citet[][C]{2000PASJ...52..113T}, and \citet{2011MNRAS.410.1774L,2015MNRAS.446.3447L,2019MNRAS.489.1533L}. For a sample of A and Am stars, \citet{1999ARep...43..819B} reported the NLTE abundances of Sr.

This paper aims to determine the NLTE abundance patterns in the range from He to Nd (for 14 elements) for a sample of nine A9 to B3 stars with low rotational velocities ($V \sin i \precsim 25$~\kms). The sample includes superficially normal as well as Am and suspected Am stars with well determined atmospheric parameters and high-quality observed spectra available.
Since 2013, our group developes and carefully tests the NLTE methods for spectroscopic analyses of A-B type stars. The results are already published for O\ione\ \citep{sitnova_o}, C\ione -\ii\ \citep{2016MNRAS.462.1123A}, Ti\ione -\ii\ \citep{sitnova_ti}, Mg\ione -\ii\ \citep{2018ApJ...866..153A}, Ca\ione -\ii\ \citep{2018MNRAS.477.3343S}, He\ione\ \citep{2018AstL...44..621K}, Si\ione -\ii -\iii\ \citep{2020MNRAS.493.6095M}, and Ne\ione\ \citep{2020ApJ...896...59A}. The work on Fe\ione -\ii\ is in progress (Sitnova 2020, in prep). In this study, we present new NLTE treatments and extend the NLTE abundance analyses to Na\ione, Sr\ii, Zr\ii -\iii, Ba\ii, and Nd\ii -\iii. 


The paper is organised as follows. Stellar sample, spectral observations, and atmospheric parameters are described briefly in Sect.~\ref{sect:obs}. Section~\ref{sect:nlte} presents new model atom for Zr\ii-\iii, updates of the model atoms for Sr\ii\ and Ba\ii, methods of the NLTE calculations, and the obtained NLTE effects for Na\ione, Sr\ii, Ba\ii, Zr\ii-\iii, and Nd\ii-\iii. In Sect.~\ref{sect:abund}, we determine abundances of O, Na, Sr, Zr, Ba, and Nd in the sample stars. Stellar element abundance patterns are discussed in Sect.~\ref{sect:pattern}. We summarise our conclusions in Sect.~\ref{sect:Conclusions}.

\section{Stellar sample, observations, atmospheric parameters}\label{sect:obs}

Our sample includes the four superficially normal stars: HD~32115 (A9~V), HD~209459 (21~Peg, B9.5~V), HD~17081 ($\pi$~Cet, B7~IV), and HD~160762 ($\iota$~Her, B3~IV), the two Am stars: HD~48915 (Sirius, A1~V) and HD~72660 (A0~V), a mild $\lambda$~Bootis-type star HD~172167 (Vega, A0~V), and the two stars: HD~73666 (40~Cnc, A1~V) and HD~145788 (A0~V), which stay separately from the others. The latter reveals an overall enhancement of metals, but was not classified as Am star \citep{2009AA...503..945F}. HD~73666 is a member of the Praesepe cluster and known as a Blue Straggler. Though \citet{2007AA...476..911F} find that HD~73666 reveals typical chemical abundances of the Praesepe cluster and refer to this star as a normal A-type star, its origin was connected, most probably, with the collisional mergers of two stars or two binary systems \citep{2010A&A...510A...8F} that can affect a chemical composition of the resulting object.
The three of our stars, namely, $\pi$~Cet, HD~32115, and HD~72660, are primary components of single line spectroscopic binaries (SB1), with negligible flux coming from the secondary star. 
Sirius is a primary component of an astrometric visual binary system. $\iota$~Her is a slowly pulsating B-type (SPB, the $\beta$~Cephei type) star. Vega is a rapidly rotating star seen pole-on. Similarly to our previous studies, we assume that it is safe to analyse the stars ignoring the presence of their secondaries for the SB1s, pulsations for $\iota$~Her, and the non-spherical effects for Vega. 

Characteristics of spectral observations are listed in Table~\ref{tab:stars_list}. For the seven stars, we used spectra observed with a spectral resolving power of $R = \lambda/\Delta\lambda >$ 60\,000 and a signal-to-noise ratio of S/N $>$ 200, using the ESPaDOnS instrument of the Canada-France-Hawaii Telescope (CHFT). For two of them, J.~Landstreet kindly provided us with the ultra-violet (UV) spectra observed on the Hubble Space Telescope with the Goddard High Resolution Spectrograph (GHRS) for Sirius \citep{1999ApJ...520..805H} and the Space Telescope Imaging Spectrograph (STIS) for HD~72660 \citep{2016MNRAS.456.3318G}.
HD~145788 was observed with the {\'{e}}chelle spectrograph HARPS (High Accuracy Radial velocity Planet Searcher) attached at the 3.6-m ESO La Silla telescope. For Vega, we used spectra observed by A. Korn with the spectrograph FOCES (fibre optics Cassegrain echelle spectrograph) at the 2.2-m telescope of the Calar Alto Observatory and by \citet{2007PASJ...59..245T} with the HIgh-Dispersion Echelle Spectrograph (HIDES) at the coud{\'{e}} focus of the 188 cm reflector at Okayama Astrophysical Observatory.

\begin{table*}
	\centering
	\caption{Atmospheric parameters of the sample stars and characteristics of the observed spectra.}
	\label{tab:stars_list}
	\begin{tabular}{lrcrccrcllr} %
		\hline\hline \noalign{\smallskip}
\multicolumn{1}{c}{Star} & $\Teff$ & $\logg$ & [Fe/H] & $\xi_t$ & Ref & $v\sin i$ & Ref &\multicolumn{3}{c}{Observed spectra} \\
\cline{9-11}
            & [K]  &      &      & [\kms ]   &  & [\kms ] & & Instrument & Source & $\lambda / \Delta \lambda$ \\
\noalign{\smallskip}\hline \noalign{\smallskip}
HD~32115    & 7250 & 4.20 & 0.00 & 2.3 & B02 & 9  & B02 & ESPaDOnS      & CAO & 60\,000 \\ 
HD~73666    & 9380 & 3.78 & 0.10 & 1.8 & F07 & 10 & F07 & ESPaDOnS      & F07     & 65\,000 \\
(40 Cnc)    &      &      &      &     &     &    &     &               &     & \\
HD~172167   & 9550 & 3.95 & $-$0.50 & 1.8 & CK93 & 23.5 & R14 & FOCES    & A. Korn & 40\,000 \\ 
(Vega)      &      &      &         &     &     &      &     & HIDES  &   T07    & 100\,000 \\
HD~72660    & 9700 & 4.10 & 0.40 & 1.8 & S16 & 7  & R14 & ESPaDOnS      & V. Khalack & 65\,000 \\
            &      &      &      &     &     &    &     & STIS          & GL16 & 110\,000 \\
HD~145788   & 9750 & 3.70 & 0.46 & 1.3 & F09 & 13 & R14 & HARPS         & F09  & 115\,000 \\
HD~48915    & 9850 & 4.30 & 0.40 & 1.8 & HL93 & 16 & HL93 & ESPaDOnS      & CAO & 65\,000 \\
(Sirius)    &      &      &      &     &     &    &     & GHRS          & W3496 & 25\,000 \\
HD~209459   & 10400 & 3.55 & 0.00 & 0.5 & F09 & 4 & S81 & ESPaDOnS      & CAO & 120\,000 \\
(21 Peg)    &       &      &      &     &     &   &     &       &     & \\
HD~17081    & 12800 & 3.75 & 0.00 & 1.0& F09 & 20 & F09 & ESPaDOnS      & F09 & 65\,000 \\
($\pi$ Cet) &      &       &      &    &     &    &     &      &     & \\
HD~160762   & 17500 & 3.80 & 0.02 & 1.0 & NP12 & 6 & NP12 & ESPaDOnS    & CAO & 65\,000 \\
($\iota$~Her) &     &      &      &     &      &   &      &         &     & \\
\noalign{\smallskip}\hline \noalign{\smallskip}
\multicolumn{11}{l}{{\bf Ref, Source:} B02 = \citet{2002A&A...389..537B}, CK93 = \citet{1993ASPC...44..496C}, } \\
\multicolumn{11}{l}{F07, F09 = \citet{2007AA...476..911F,2009AA...503..945F}, GL16 = \citet{2016MNRAS.456.3318G},} \\
\multicolumn{11}{l}{HL93 = \citet{1993AA...276..142H}, NP12 = \citet{2012A&A...539A.143N}, R14 = \citet{2014A&A...562A..84R}, } \\
\multicolumn{11}{l}{S81 = \citet{1981PASP...93..587S}, S16 = \citet{sitnova_ti},} \\
\multicolumn{11}{l}{T07 = \citet[][{\tt http://pasj.asj.or.jp/v59/n1/590122/590122-frame.html}]{2007PASJ...59..245T}, } \\
\multicolumn{11}{l}{W3496 = GO Proposal 3496 by G. Wahlgren, CAO (Common Archive Observation database) = } \\
\multicolumn{11}{l}{ {\tt http://www.cfht.hawaii.edu/Instruments/Spectroscopy/Espadons/}, } 
	\end{tabular}
\end{table*}

 

For consistency with our earlier studies, for each star we used exactly the same effective temperature ($\Teff$), surface gravity ($\logg$), metallicity ([Fe/H]), and microturbulent velocity ($\xi_t$). The data and its sources are listed in Table~\ref{tab:stars_list}. 
 We refer to the original papers and also \citet{2016MNRAS.462.1123A,2018ApJ...866..153A}, and \citet{sitnova_ti} for a description of the methods of atmospheric parameter determinations. 

Classical plane-parallel and LTE model atmospheres were calculated with the code \textsc{LLmodels} \citep{2004AA...428..993S}. For Sirius, its model atmosphere was taken from R.~Kurucz website\footnote{http://kurucz.harvard.edu/stars/sirius/ap04t9850g43k0he05y.dat}.

\section{Non-LTE calculations}\label{sect:nlte}

\subsection{New model atom of Zr\ii-\iii }

We produced a new model atom, which includes the model atom of Zr\ii, as constructed by \citet{Velichko2010_zr}, and the Zr\iii\ states with an excitation energy of \Eexc\ $\precsim$ 12.2~eV (up to \eu{5s5p}{1}{P}{\circ}{}) from measurements of \citet{1997PhyS...55..310R}, as provided by the National Institute of Standards and Technology (NIST) database\footnote{https://physics.nist.gov/PhysRefData/ASD} \citep{NIST19}. The triplet fine structure was taken into account for the Zr\iii\  \eu{4d^2}{3}{F}{}{} and \eu{4d^2}{3}{P}{}{} terms. The Zr\iii\ levels above 12.2~eV were ignored because, in the stellar parameter range with which we concern, they do not affect the statistical equilibrium (SE) of zirconium and contribute less than 0.0001\%\ to the total abundance of Zr. Final model atom includes 247 levels of Zr\ii, 18 levels of Zr\iii, and the ground state of Zr\iv. 

Oscillator strengths predicted by R.~Kurucz\footnote{http://kurucz.harvard.edu/atoms/} were used to compute radiative rates for 126 transitions of Zr\iii. We compared the adopted $gf$-values with predictions of \citet{2020A&A...637A..10R} for 112 transitions in common. 
\citet{2020A&A...637A..10R} provide slightly lower $gf$-values compared with that of R.~Kurucz, by $0.06\pm0.26$~dex for all the transitions and by $0.09\pm0.11$~dex for 68 transitions with $\log gf > -1.5$. We note that \citet[][see their Fig.~5]{2020A&A...637A..10R} underestimate the transition probabilities compared with laboratory measurements of \citet{2005JQSRT..94..109M}, by 0.07~dex, on average. Thus, our adopted $gf$-values agree very well the experimental data.

For both Zr\ii\ and Zr\iii, we did not find in the literature accurate data on photoionization and electron-impact excitation cross sections. The photoionization cross sections were calculated in the hydrogenic approximation, where a principal quantum number $n$ was replaced with an effective principal quantum number $n_{\rm eff}$. The formula of \citet{Reg1962} for allowed transitions and $\Upsilon$ = 1 for forbidden transitions were applied to obtain the collisional rates.
Ionization by electronic collisions was calculated from the \citet{1962amp..conf..375S} formula, employing a hydrogenic photoionization cross-section at the threshold.


\begin{figure*}           
 \begin{minipage}{150mm}

\hspace{-10mm}
\parbox{0.3\linewidth}{\includegraphics[scale=0.35]{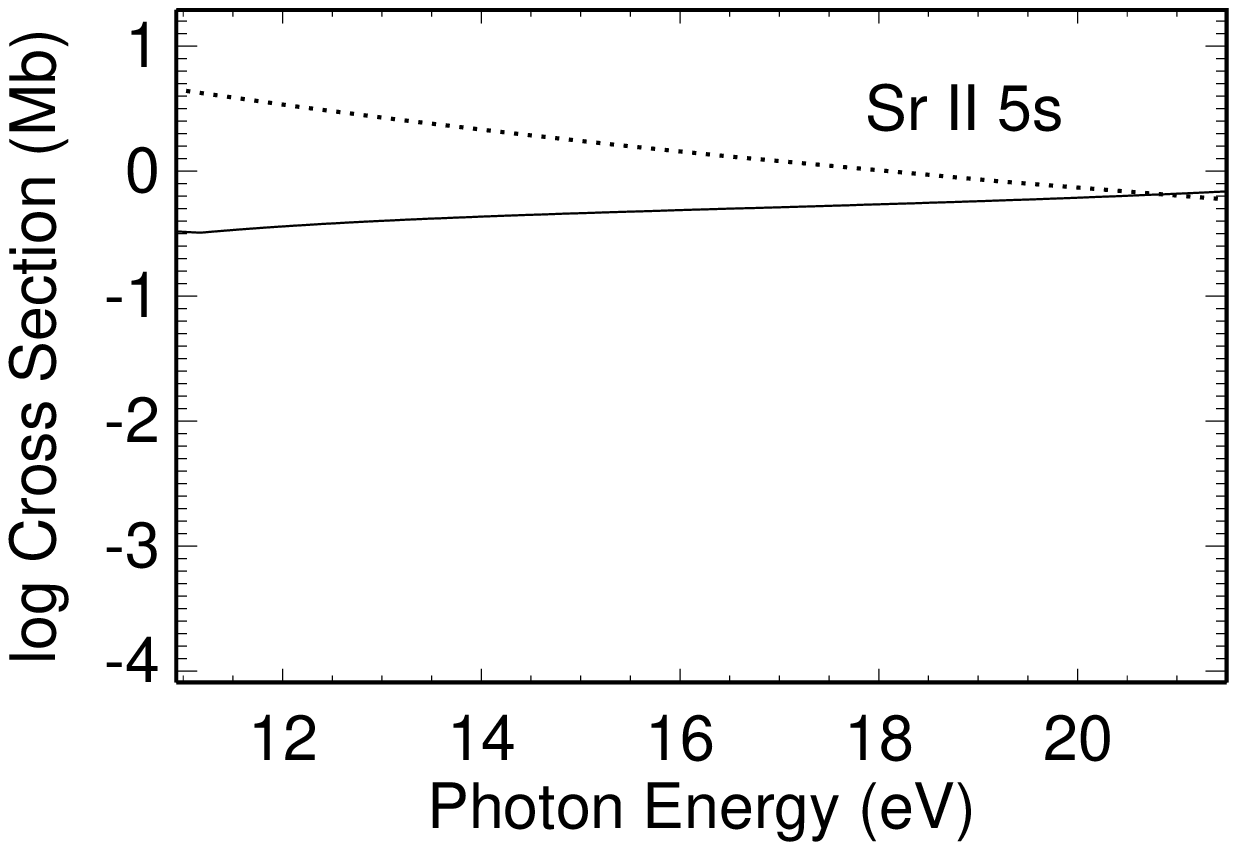}\\
\centering}
\hspace{5mm}
\parbox{0.3\linewidth}{\includegraphics[scale=0.35]{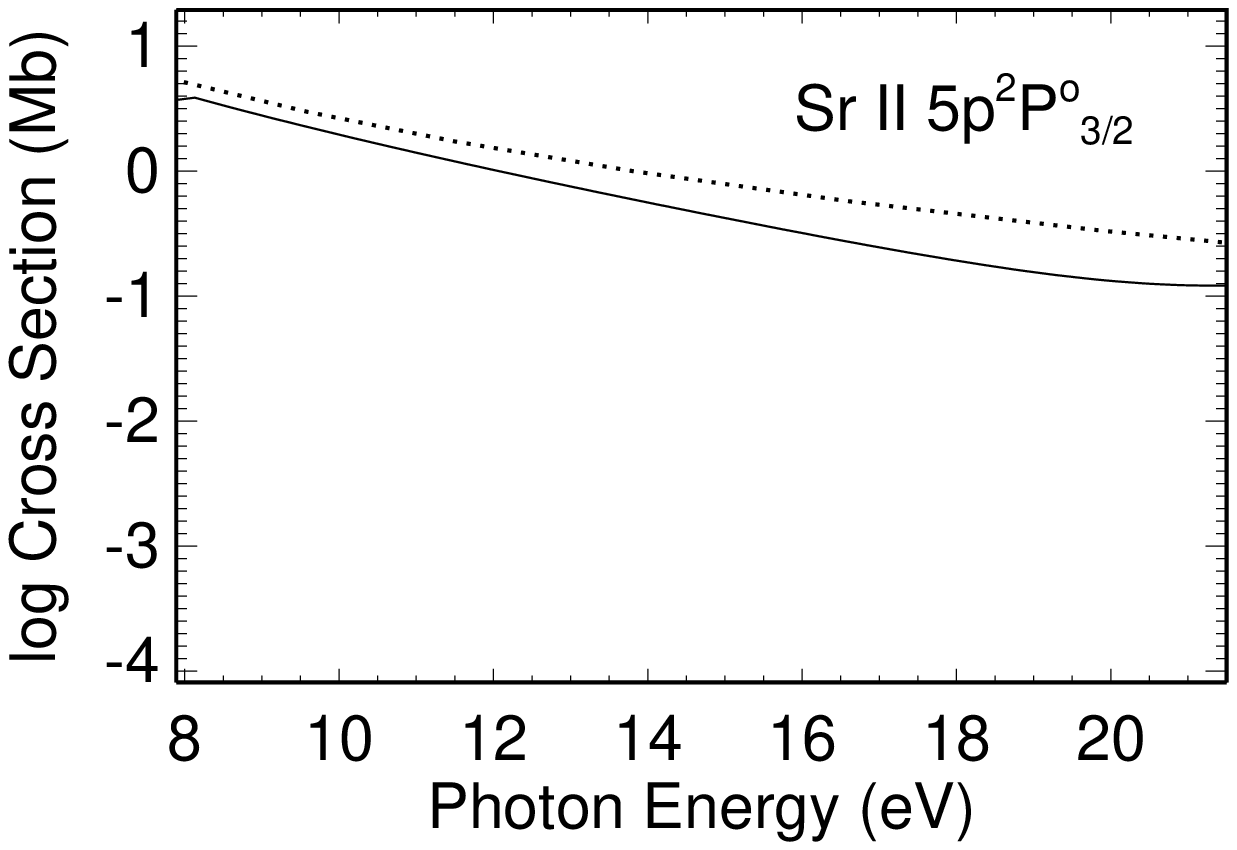}\\
\centering}
\hspace{5mm}
\parbox{0.3\linewidth}{\includegraphics[scale=0.35]{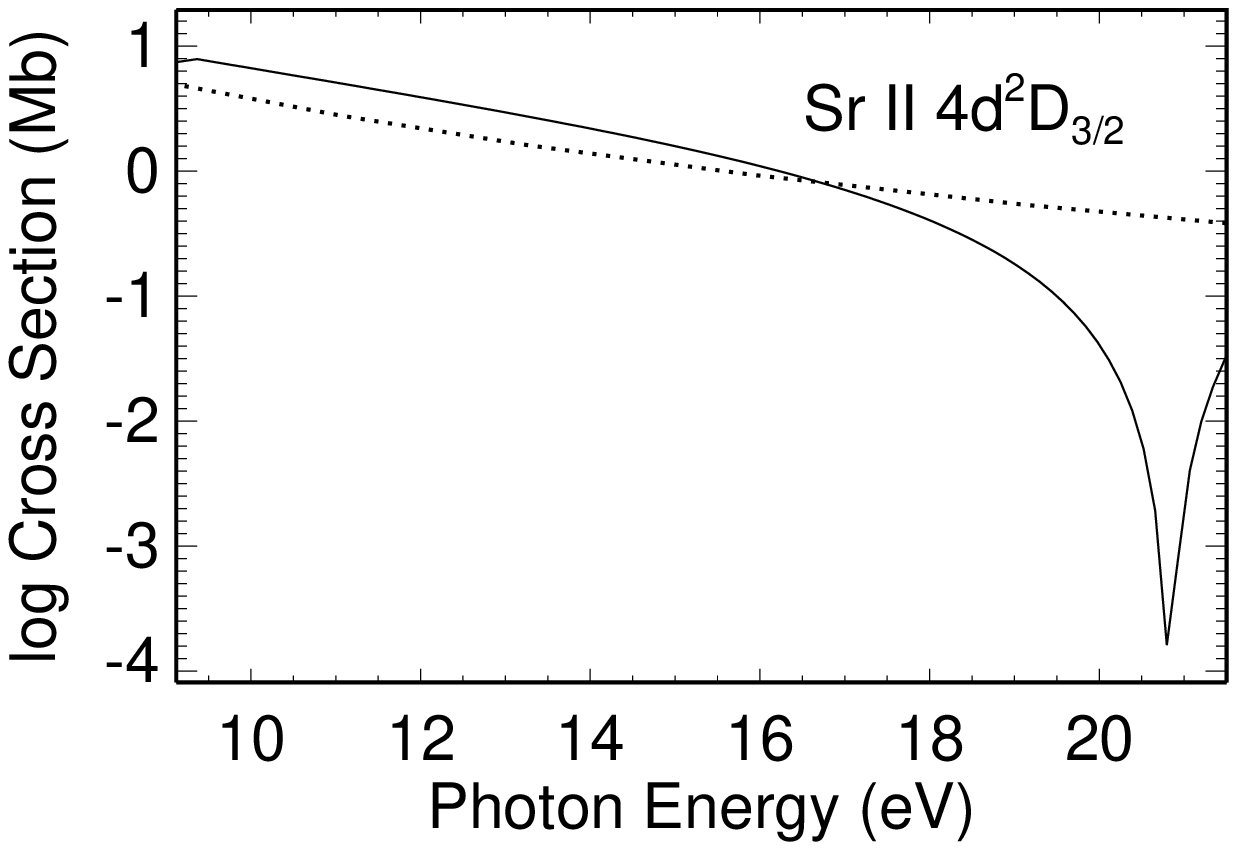}\\
\centering}
\hfill
\\[0ex]

\vspace{-7mm}
\hspace{-10mm}
\parbox{0.3\linewidth}{\includegraphics[scale=0.35]{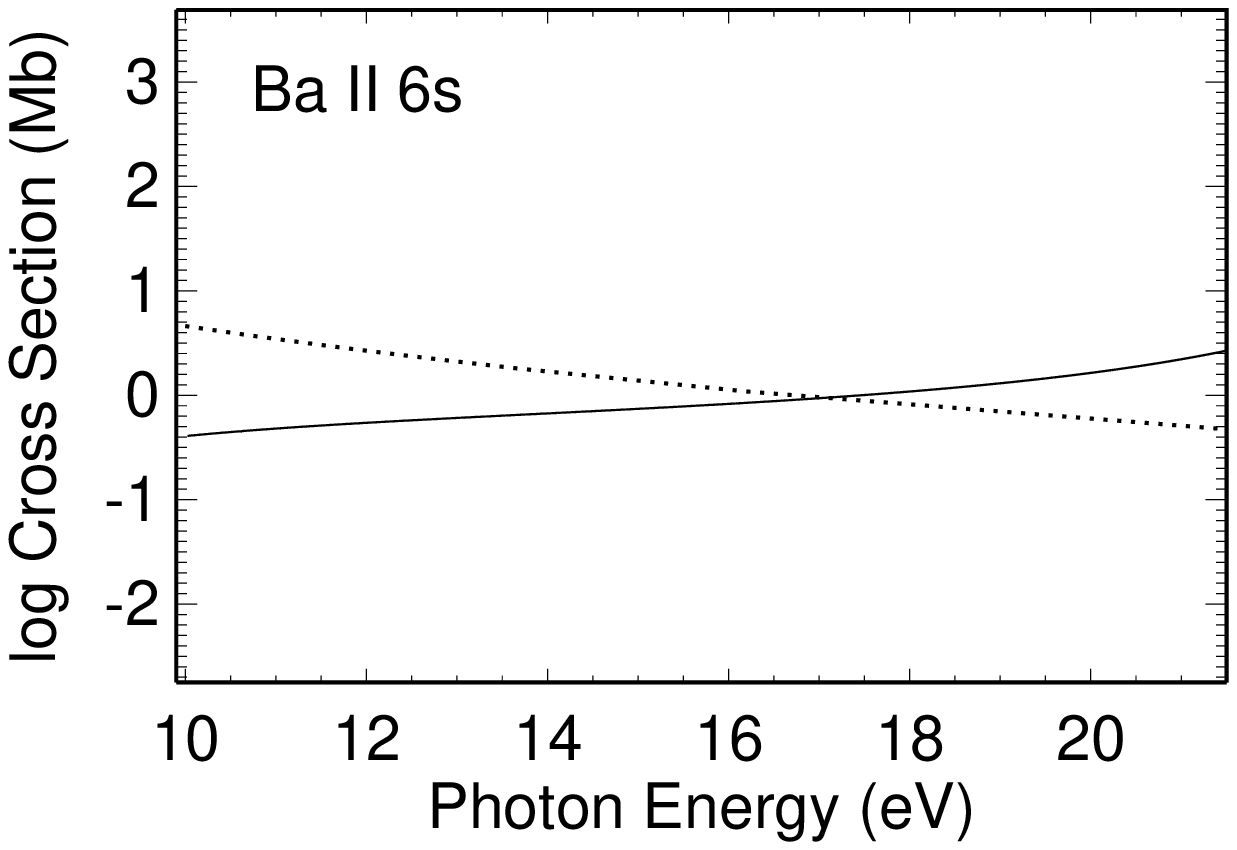}\\
\centering}
\hspace{5mm}
\parbox{0.3\linewidth}{\includegraphics[scale=0.35]{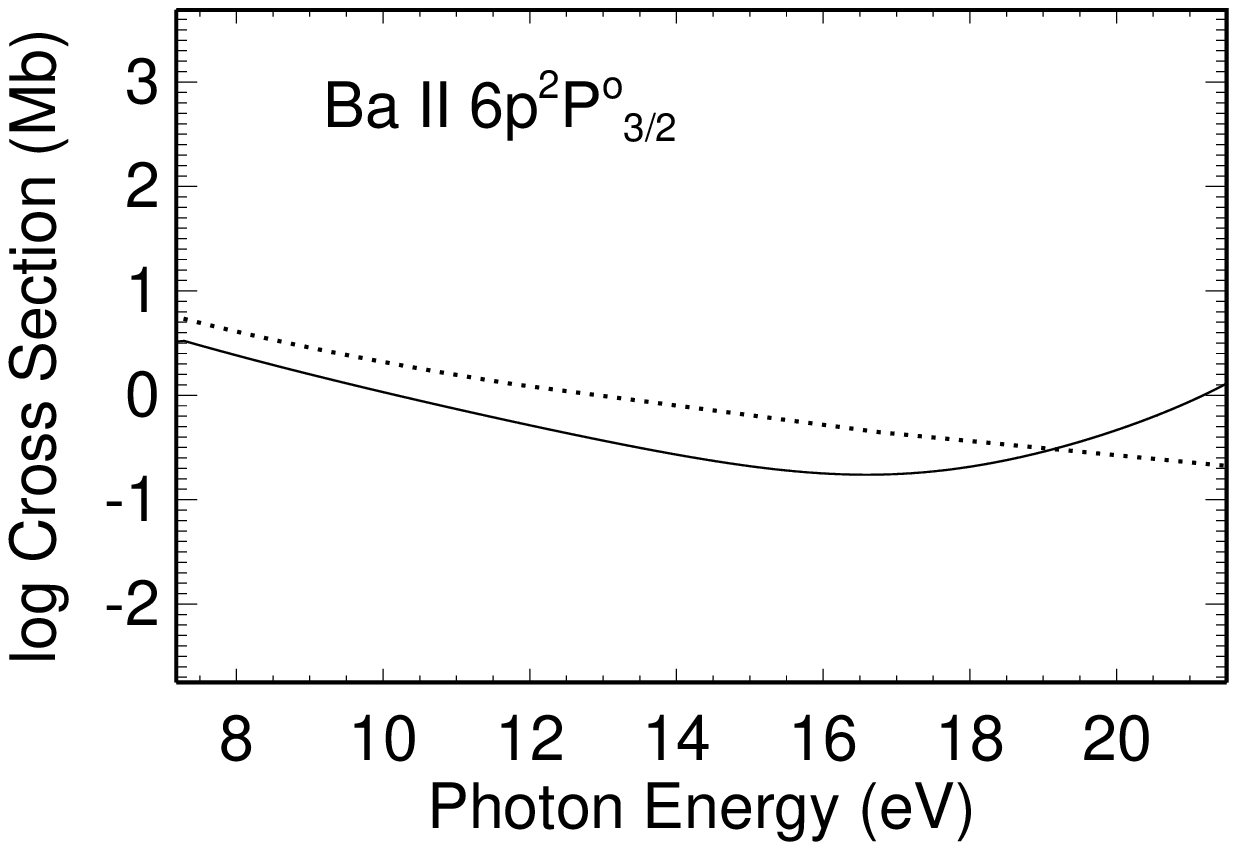}\\
\centering}
\hspace{5mm}
\parbox{0.3\linewidth}{\includegraphics[scale=0.35]{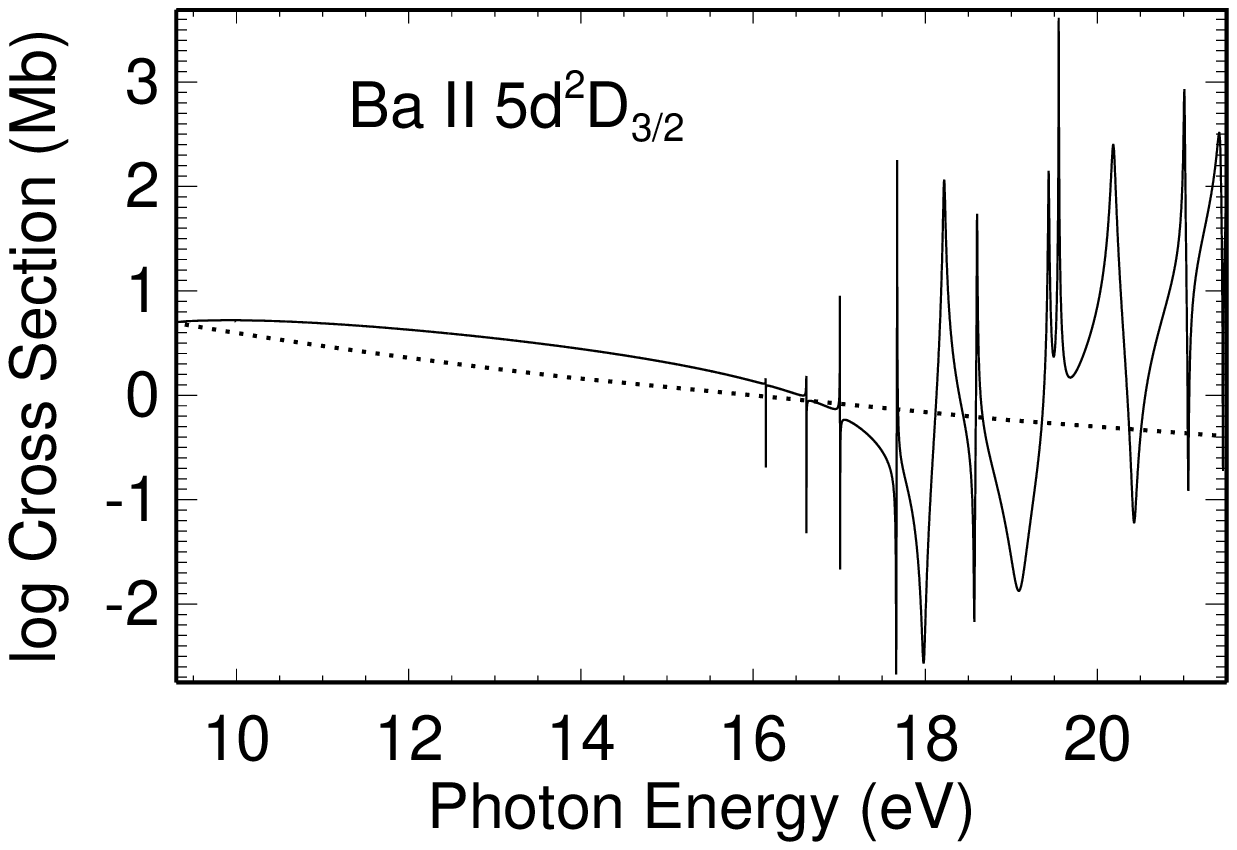}\\
\centering}
\hfill
\\[0ex]

\vspace{-7mm}
\caption{Photoionization cross sections for the low-excitation levels of Sr\ii\ (top row) and Ba\ii\ (bottom row) as a function of photon energy from the calculations with the R-matrix method (solid curves) and in hydrogenic approximation (dotted curves). 
}
\label{fig:photo}
\end{minipage}
\end{figure*} 

\subsection{Update of the model atom for Sr\ii }\label{sect:sr2}

A comprehensive model atom of Sr\ii\ was originally constructed by \citet{1997ARep...41..530B} and updated in this study, as follows.

More Sr\ii\ energy levels, up to $n$ = 14, with the data from NIST, were included in the atomic model. The doublet fine structure is taken into account for the Sr\ii\ \eu{4d}{2}{D}{}{} and \eu{5p}{2}{P}{\circ}{} terms. Sr\iii\ is represented by the ground state because the excited states, with their excitation energies of \Eexc\ $> 21.8$~eV, contribute only a little to the Sr abundance, even in the hot atmospheres. Oscillator strengths of the newly added transitions were taken from calculations of R.~Kurucz.

In the atmospheres with $\Teff >$ 9000~K, Sr\ii\ is strongly ionized, and its SE depends strongly on deviations of the mean intensity of ionizing radiation from the Planck function and the photoionization cross sections. 
For the Sr\ii\ states from 5s to 8s, the photoionization cross sections were calculated in this study with the Dirac $B$-spline $R$-matrix (DBSR) method, in the same approximations as it was done for h$\nu$ + K\ione\ problem \citep{2010PhRvA..81d3423Z}. New data for the three lowest terms are displayed in Fig.~\ref{fig:photo}. The hydrogenic approximation (with $n_{\rm eff}$) 
was used for the remaining levels in the model atom.

For all the transitions between 5s and up to 4f, we apply electron-impact excitation rate coefficients from ab initio calculations of \citet{2002MNRAS.331..875B}. The formula of \citet{Reg1962} for allowed transitions and the effective collision strength $\Upsilon$ = 1 for forbidden transitions were used in the other cases. Ionization by electronic collisions was calculated as for Zr\ii -\iii.

\subsection{Update of the model atom for Ba\ii }

Similarly to Sr\ii, Ba\ii\ is a minority species in the atmospheres with $\Teff >$ 9000~K. Therefore, the system of energy levels has to be fairly complete in the SE calculations. The model atom constructed by \citet{Mashonkina1999} was extended by including all the  Ba\ii\ levels up to $n$ = 50, as provided by NIST \citep{NIST19}. The doublet fine structure was taken into account for the \eu{5d}{2}{D}{}{} and \eu{6p}{2}{P}{\circ}{} terms. Only the ground state of Ba\iii\ was included because the excited states lie above \Eexc\ = 16.5~eV and do not affect the SE of barium. For the transitions missing in model atom of \citet{Mashonkina1999}, their radiative rates were computed with oscillator strengths predicted by R.~Kurucz.

The photoionization cross sections for the Ba\ii\ states from 6s to 6f were calculated in this study with the DBSR method, following \citet{2010PhRvA..81d3423Z}. Figure~\ref{fig:photo} displays new data for the three lowest terms. For the remaining levels in the model atom, we applied the hydrogenic approximation with $n_{\rm eff}$.

Experimental data of \citet{1974PhRvA..10..141C} are available for electron-impact excitation of the Ba\ii\ 6p, 7s, and 6d levels from the ground state. For all the remaining transitions between 6s and 6d, we applied the collision cross sections computed by \citet{1981eabs.book.....S} in the Born approximation. If the data is absent the formula of \citet{Reg1962} for allowed transitions and $\Upsilon$ = 1 for forbidden transitions have been adopted. Ionization by electronic collisions was calculated as for Zr\ii -\iii.

\subsection{Oxygen, sodium, neodimium}

The  O\ione\ model atom of \citet{2000A&A...359.1085P} and \citet{sitnova_o} was updated, by calculating radiative rates of strong UV bound-bound (b-b) transitions with the Voigt absorption profiles. With this model atom, we revised the NLTE abundances of oxygen in HD~32115, Vega, and Sirius compared with those of \citet{sitnova_o} and determined the O abundances of the remaining stellar sample. 

Non-LTE calculations for Na\ione\ were performed with the model atom of \citet{alexeeva_na} and for Nd\ii -\iii, using the model atom of \citet{2005A&A...441..309M}.

\subsection{Codes, list of investigated lines}

All the NLTE species were treated as the trace ones assuming that atmospheric structure is not affected by deviations of their level populations from the thermodynamic equilibrium ones and the SE calculations can be performed with fixed LTE model atmosphere. \citet{2011JPhCS.328a2015P} showed that such a hybrid method 
is able to reproduce observations for $\Teff$ = 15\,000 to 35\,000~K. This is all the more true for our cooler atmospheres. 
We employed a modified version of the {\sc DETAIL} code \citep{detail,2011JPhCS.328a2015P} to solve the coupled radiative transfer and SE equations.  

Spectral lines used in abundance analyses, together with their atomic parameters, are listed in Table~\ref{tab:linelist}. 

In the visual and near infrared (IR) spectral range of most of our stars with $\Teff \le$ 10400~K, sodium and barium are well represented, by up to eight lines of Na\ione\ and five lines of Ba\ii, while strontium, zirconium, and neodimium are observed in two lines, at most. The exception is HD~72660, for which we measured six lines of Zr\iii\ in the UV spectrum. For lines of Na\ione\ and Sr\ii, we used $gf$-values recommended by NIST. Laboratory $gf$-values from \citet{2015PhRvA..91d0501D} and \citet{2016NatSR...629772D} were adopted for lines of Ba\ii\ and from \citet{LNAJ} for Zr\ii. For Zr\iii\ and Nd\iii, $gf$-values are provided by the Vienna Atomic Line Database \citep[VALD,][]{2015PhyS...90e4005R}. Hyperfine splitting (HFS) and/or isotopic components were taken into account for lines of Ba\ii\ and Sr\ii, using the data from \citet{1998AJ....115.1640M} and \citet{1983HyInt..15..177B}, respectively. The fractional isotope abundances correspond to the solar system matter \citep{Lodders2009}.

 For lines of Na\ione\ and Ba\ii, quadratic Stark effect broadening was treated using the full width at half maximum intensity (FWHM) from the STARK-B\footnote{https://stark-b.obspm.fr} database.
For the Sr\ii\ resonance lines, log~$\Gamma_4 = -5.40$ was computed by E.~Yukov, as published by \citet{1975Ap&SS..34..403K}.
 For the remaining lines, the $\Gamma_4$ values were computed from the approximate formula of \citet{1971Obs....91..139C}. Van der Waals broadening is important only for the coolest star of our sample, HD~32115. We applied accurate $\Gamma_6$ values from \citet{BPM} for lines Na\ione, Sr\ii, and Ba\ii. For the remaining lines van der Waals broadening was taken into account with the \citet{1955psmb.book.....U} formula.

 We also included the Sc\ii\ lines in our analysis because the Sc abundance is often used as one of the characteristics in Am-star classification.
Abundances of Sc were determined under the LTE assumption. Expected NLTE effects are discussed in Sect.~\ref{sect:scandium}. Laboratory $gf$-values of the Sc\ii\ lines were taken from \citet{1989JOSAB...6.1457L} and the HFS components were calculated by \citet{2019ARep...63.1010P} using experimental data of \citet{MDYC} and \citet{VLAH}.


Abundances from individual lines were derived from  
line profile fitting, but not from equivalent widths ($EW$s). 
Exceptions are Vega, where the line profiles have a rectangular shape, and the LTE abundances from the O\ione\ 7771, 7774, 7775\,\AA\ lines, whose profiles in each star cannot be reproduced under the LTE assumption.
In order to compute the synthetic spectra, we used the {\sc Synth}V\_NLTE code \citep{2019ASPC}, which implements the pre-computed departure coefficients from {\sc DETAIL}. The line list required for calculations of the synthetic spectra, together with atomic data, was taken from VALD.
The {\sc IDL binmag} code by O. Kochukhov\footnote{http://www.astro.uu.se/$\sim$oleg/binmag.html}  was used to obtain the best fit to the observed spectrum.  
An $EW$ analysis was made also with the {\sc Synth}V\_NLTE + {\sc IDL binmag} code. 


\begin{figure*}
 \begin{minipage}{170mm}

\hspace{-5mm}
\parbox{0.5\linewidth}{\includegraphics[scale=0.6]{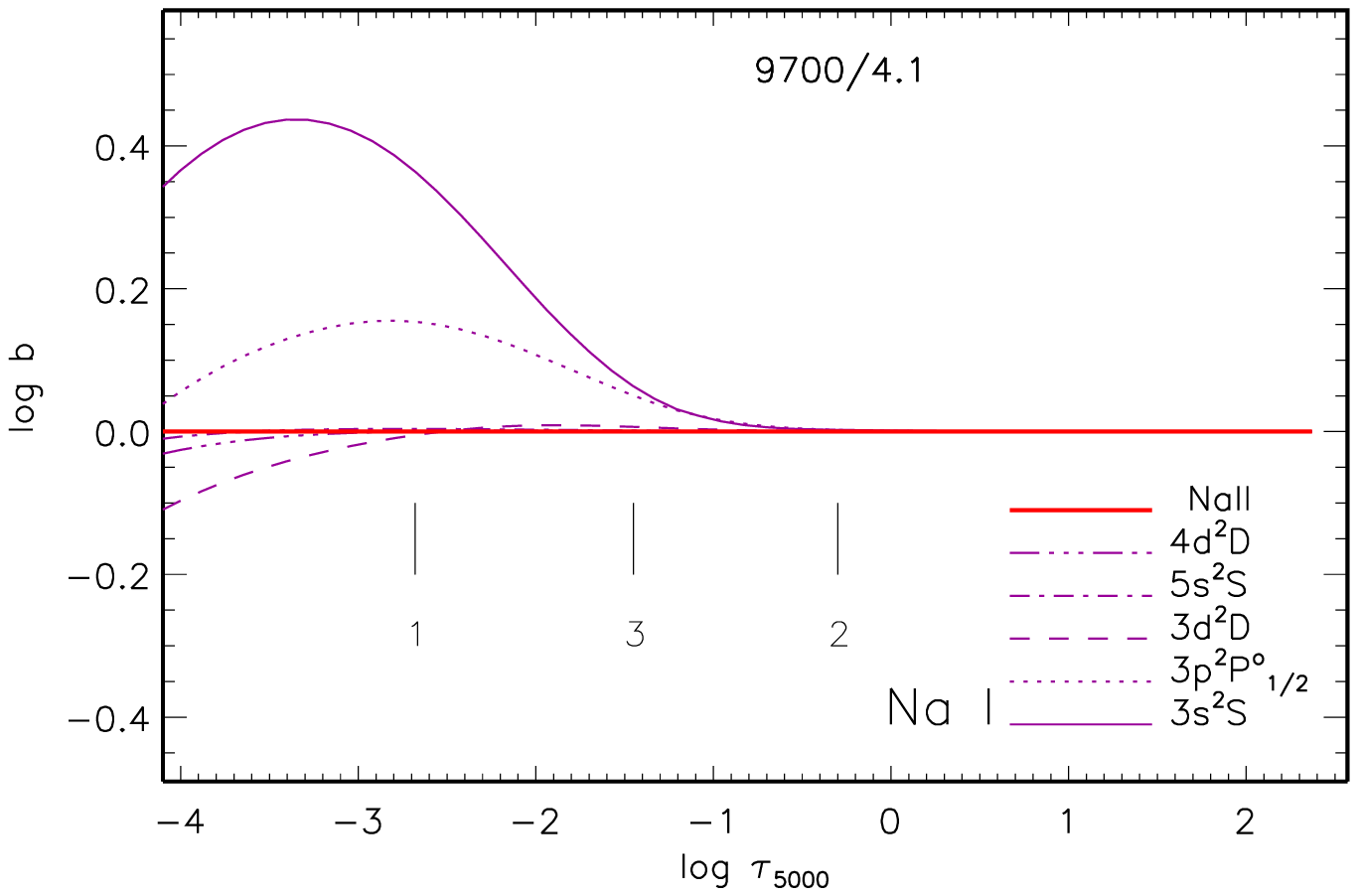}\\
\centering}
\hspace{5mm}
\parbox{0.5\linewidth}{\includegraphics[scale=0.6]{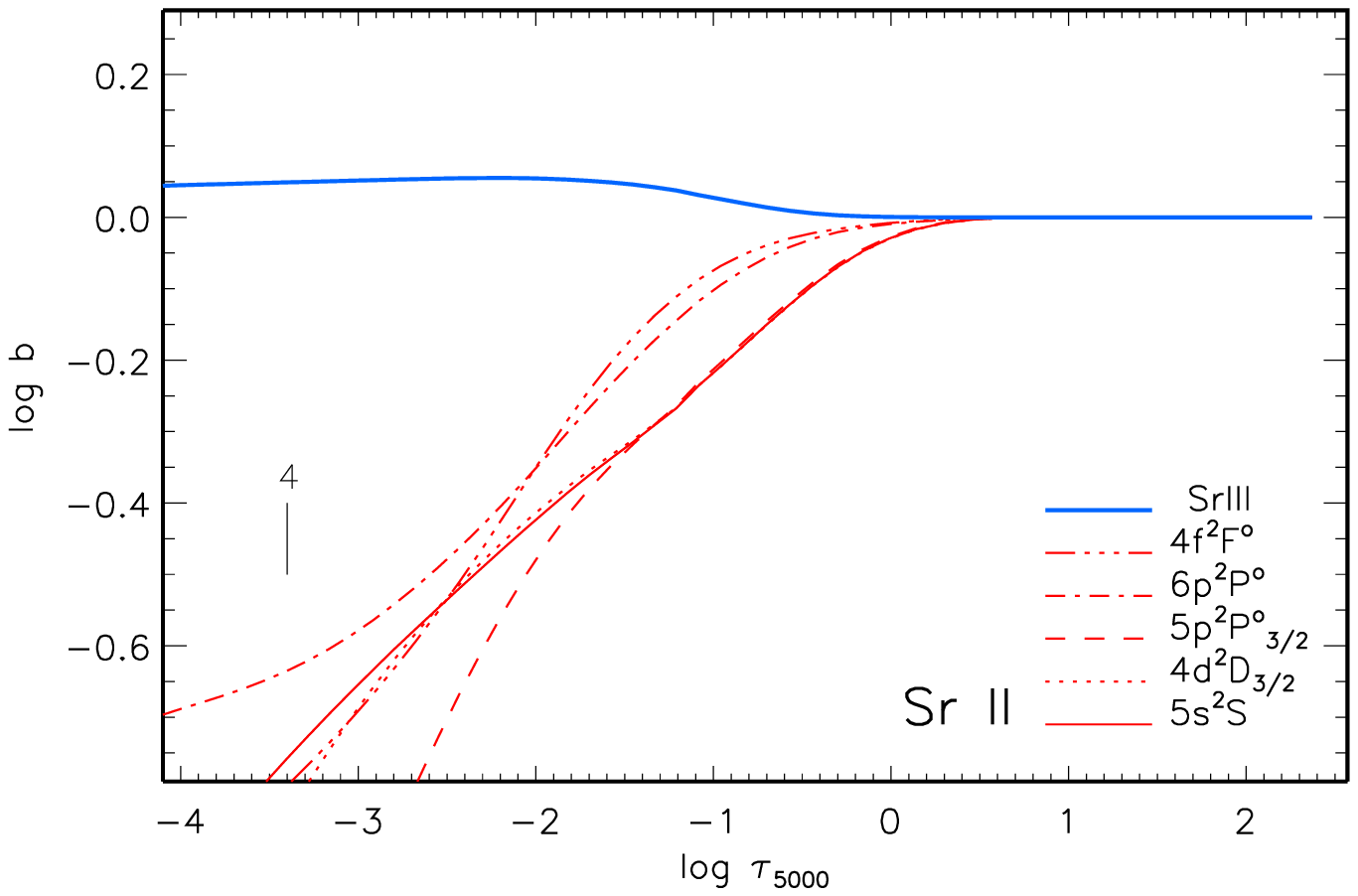}\\
\centering}
\hfill
\\[0ex]

\vspace{-7mm}
\hspace{-5mm}
\parbox{0.5\linewidth}{\includegraphics[scale=0.6]{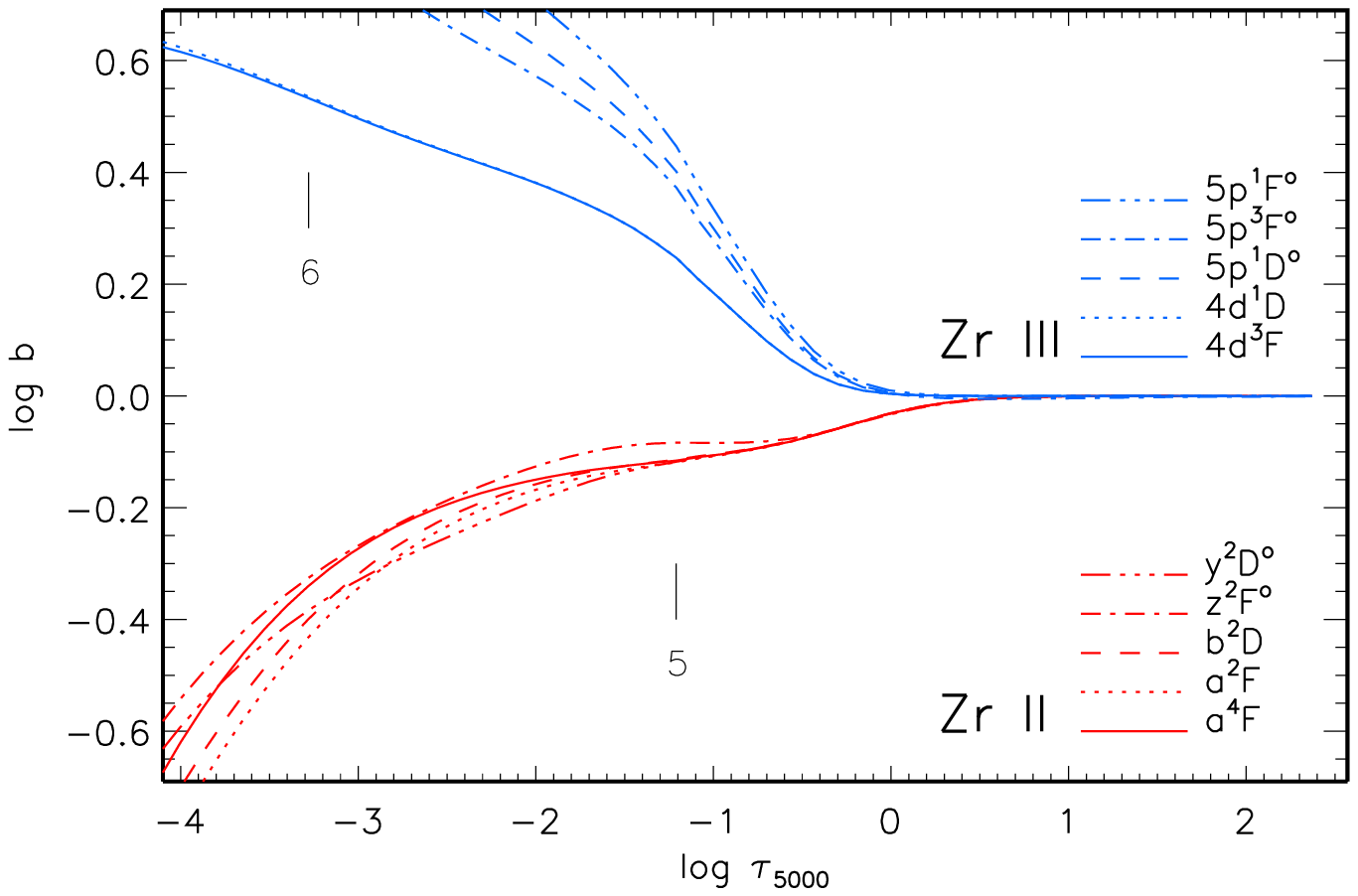}\\
\centering}
\hspace{5mm}
\parbox{0.5\linewidth}{\includegraphics[scale=0.6]{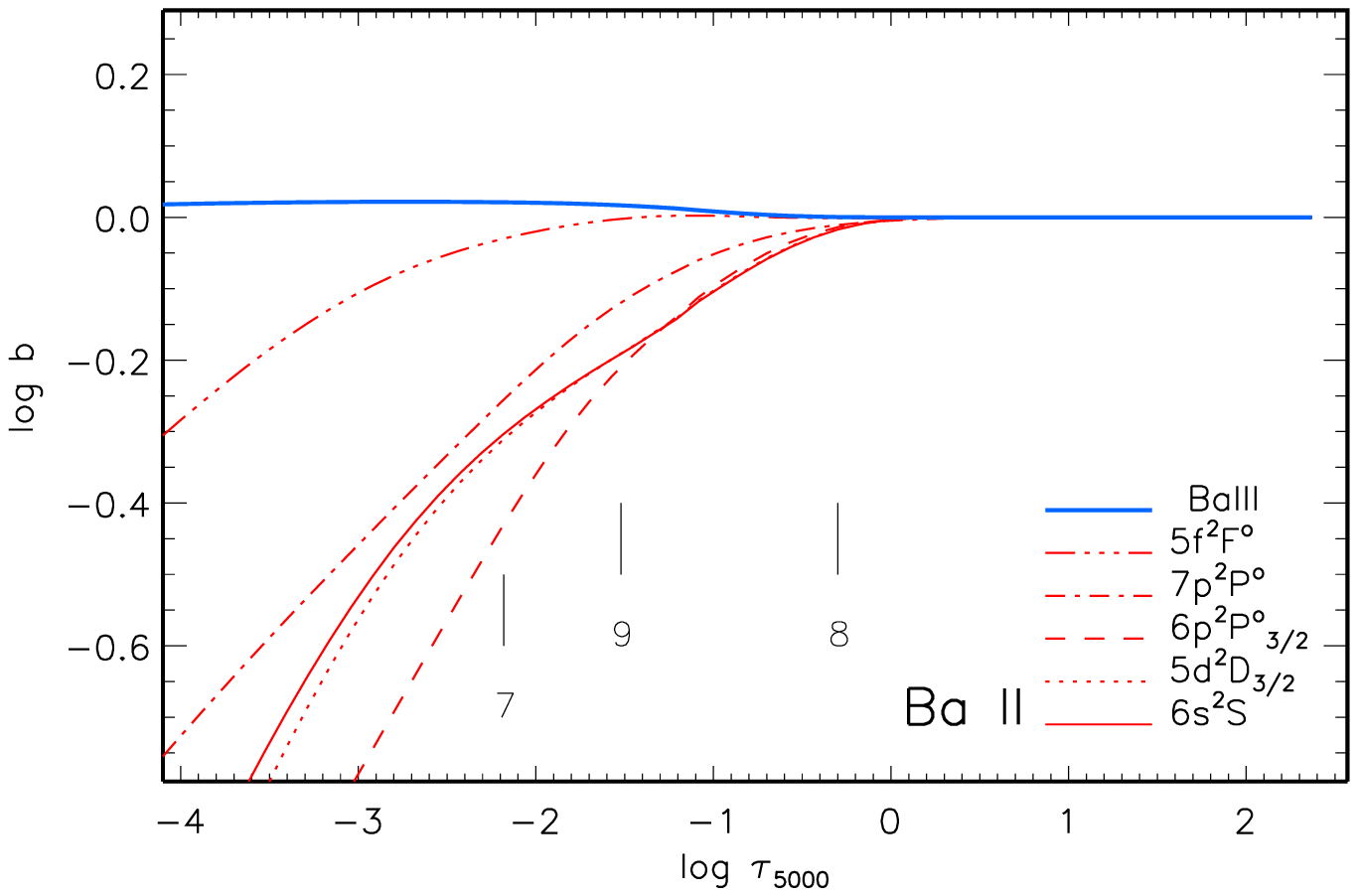}\\
\centering}
\hfill
\\[0ex]

\vspace{-7mm}
    \caption{Departure coefficients, $b$, for the levels of Na\ione\ (top left panel), Sr\ii\ (top right panel), Zr\ii -Zr\iii\ (bottom left panel), and Ba\ii\ (bottom right panel) as a function of $\log \tau_{5000}$ in the model atmosphere 9700/4.10/0.4. The selected levels are quoted in the right part of each panel. Tick marks indicate the locations of line center optical depth unity for the following lines: Na\ione\ 5889 (1), 5688 (2), and 8194\,\AA\ (3), Sr\ii\ 4077\,\AA\ (4), Zr\ii\ 4149\,\AA\ (5), Zr\iii\ 1790\,\AA\ (6), and Ba\ii\ 4554 (7), 5853 (8), and 6141\,\AA\ (9). }
    \label{fig:bf}
\end{minipage}
\end{figure*}

\subsection{Non-LTE effects}

The NLTE effects for C\ione -\ii, O\ione, Ne\ione, Mg\ione -\ii, Si\ione -\ii -\iii, Ca\ione -\ii, and Ti\ione -\ii\ were discussed in detail in our previous papers. Here, we concentrate  on the new NLTE species. For each of them, a behavior of the departure coefficients, ${\rm b = n_{NLTE}/n_{LTE}}$, is similar in the model atmospheres with $\Teff >$ 9000~K. Here, ${\rm n_{NLTE}}$ and ${\rm n_{LTE}}$ are the statistical equilibrium and thermal (Saha-Boltzmann) number densities, respectively. We select the model with $\Teff$/$\logg$/[Fe/H] = 9700/4.10/0.4 to display log~$b$ for the selected levels of Na\ione, Sr\ii, Zr\ii -Zr\iii, and Ba\ii\ in Fig.~\ref{fig:bf}.  

\subsubsection{Na\ione}
NLTE leads to an overpopulation of the  Na\ione\ ground (3s) and first excited (3p) levels due to the cascade from upper levels induced by the escape of line photons \citep[photon-suction effect, as called by][]{1992A&A...265..237B}. Lines of Na\ione\ are strengthened, and the NLTE abundance corrections, 
 $\Delta_{\rm NLTE} = \eps{NLTE} - \eps{LTE}$, are negative.
The NLTE effects are stronger for the 5889 and 5895\,\AA\ resonance lines compared with the subordinate lines, and, for the subordinate lines arising from common 3p level, they are stronger for the stronger lines, which form in the higher atmospheric layers. For example, in the 9700/4.10/0.4 model, $\Delta_{\rm NLTE}$ = $-0.77$, $-0.47$, and $-0.10$~dex for Na\ione\ 5889, 8194, and 5688\,\AA, respectively. Magnitude of $\Delta_{\rm NLTE}$ depends weakly on $\Teff$.

\subsubsection{Sr\ii\ and Ba\ii}
For both species, the NLTE mechanism depends on $\Teff$. In the atmosphere of our coolest star, HD~32115, Sr\ii\ and Ba\ii\ are the majority species, and the NLTE effects are mostly determined by dropping the line source function ($S_\nu$) below the Planck function ($B_\nu$) in the line-formation layers. Lines of Sr\ii\ and Ba\ii\ are strengthened, and the NLTE abundance corrections are negative, with $\Delta_{\rm NLTE}$ = $-0.04$, $-0.09$, and $-0.21$~dex for Sr\ii\ 4077\,\AA, Ba\ii\ 4554, and 6496\,\AA, respectively. 

In hot atmospheres, with $\Teff >$ 9000~K, Sr\ii\ and Ba\ii\ are strongly ionized, and the main NLTE mechanism is the overionization caused by superthermal radiation of a non-local origin below the thresholds of the $4d$ and $5p$ levels for Sr\ii\ and the $5d$ and $6p$ levels for Ba\ii. A depletion of the level populations in the line-formation layers (Fig.~\ref{fig:bf}) results in weakened lines and positive NLTE abundance corrections, which 
grow toward higher $\Teff$. For example, $\Delta_{\rm NLTE}$ = 0.35 and 0.26~dex for Sr\ii\ 4077\,\AA\ and Ba\ii\ 4554\,\AA\ in the 9380/3.78/0.10 model, and the corresponding numbers are 0.59 and 0.38~dex in the 10400/3.55/0.0 model.

\subsubsection{Zr\ii -\iii}
Similarly to Sr\ii\ and Ba\ii, Zr\ii\ moves from a majority species in the 7250/4.20/0.0 model to a minority one for $\Teff >$ 9000~K. However, Zr\ii\ has different atomic term structure compared with that of Sr\ii\ and Ba\ii, such that NLTE leads to slightly weakened lines of Zr\ii\ in our coolest model atmosphere, with $\Delta_{\rm NLTE}$ = 0.05 to 0.08~dex for different lines. For $\Teff >$ 9000~K, Zr\ii\ is subject to overionization and $\Delta_{\rm NLTE}$ grows toward higher $\Teff$. For example, for Zr\ii\ 4149\,\AA, $\Delta_{\rm NLTE}$ = 0.19 and 0.51~dex in the 9380/3.78/0.10 and 10400/3.55/0.0 models, respectively. 

Thanks to relatively high ionization energy of Zr\ii\ ($\chi_{\rm ion} \simeq$ 13.1~eV), Zr\iii\ and Zr\ii\ have comparable number densities in the atmospheres with $\Teff$ = 9700 and 9850~K. Therefore, overionization of Zr\ii\ in the line-formation layers leads to enhanced populations of the Zr\iii\ levels (Fig.~\ref{fig:bf}) and strengthened lines of Zr\iii, with negative NLTE abundance corrections. In the 9700/4.10/0.40 model, $\Delta_{\rm NLTE}$ = $-0.10$ to $-0.17$~dex for different lines.

\subsubsection{Nd\ii -\iii}
Lines of Nd\ii\ were measured only in our coolest star, HD~32115. In its atmosphere, Nd\ii\ is a majority species and NLTE leads to slightly weakened lines, with $\Delta_{\rm NLTE}$ = 0.12~dex for both Nd\ii\ 4706 and 5319\,\AA.

Nd\iii\ is a majority species in the atmospheres with $\Teff >$ 9000~K. The NLTE effects are minor, with $\Delta_{\rm NLTE} \le$ 0.01~dex in absolute value for the lines at 5294 and 5203\,\AA, which arise from the ground term.

\subsection{Comparison with other NLTE studies}

In the stellar parameter range, with which we concern, the NLTE calculations for Na\ione\ were performed by \citet{2009PASJ...61.1165T}. For Na\ione\ 5889 and 5895\,\AA\ in Sirius, with slightly higher $\Teff$ = 9938~K compared with ours, by 90~K, \citet{2009PASJ...61.1165T} report the average $\Delta_{\rm NLTE}$ = $-0.67$~dex. We could not measure Na\ione\ 5889\,\AA\ in Sirius and obtained $\Delta_{\rm NLTE}$ = $-0.56$~dex for the second resonance line. We note that, in all the cases where both lines were measured, we calculated greater $\Delta_{\rm NLTE}$ for Na\ione\ 5889\,\AA\ compared with that for Na\ione\ 5895\,\AA. For example, $\Delta_{\rm NLTE}$ = $-0.77$ and $-0.54$~dex for Na\ione\ 5889 and 5895\,\AA, respectively, in the 9700/4.10/0.40 model.
                      
\citet{Gigas88} perform the NLTE calculations for Ba\ii\ in Vega and report 
$\Delta_{\rm NLTE}$ = 0.30 and 0.27~dex for Ba\ii\ 4554 and 4934\,\AA, respectively. We obtain larger NLTE abundance corrections, by 0.09 and 0.10~dex.

\begin{table*}
	\centering
	\caption{Line atomic data and LTE (L) and NLTE (N) abundances, $\eps{}$, from individual lines in the sample stars.  }
	\label{tab:linelist}
	\begin{tabular}{lrrcccccccccccccccc} %
		\hline\hline \noalign{\smallskip}
Atom & \Eexc & log $gf$ & $\log \Gamma_4/N_{\rm e}$ & $\log \Gamma_6/N_{\rm H}$ & \multicolumn{2}{c}{HD~32115} & \multicolumn{2}{c}{...}  &  \multicolumn{2}{c}{HD~72660}  & \multicolumn{2}{c}{...} &  \multicolumn{2}{c}{21 Peg}   & \multicolumn{2}{c}{...} &\multicolumn{2}{c}{$\iota$ Her}   \\
     &  &  &  &   &  L   & N    &  \multicolumn{2}{c}{...} &   L   & N    & \multicolumn{2}{c}{...}  &   L   & N    &\multicolumn{2}{c}{...} &  L   & N     \\
		\hline \noalign{\smallskip}
\multicolumn{19}{l}{O\ione} \\
7771.94 & 9.15 &  0.37 & -5.55 &  -7.47 &  9.69 & 9.01 & \multicolumn{2}{c}{...} &  9.65 & 8.49 & \multicolumn{2}{c}{...} & 10.14 & 8.60 & \multicolumn{2}{c}{...}  & 10.35 & 8.68 \\
\multicolumn{19}{l}{Na\ione} \\
5682.63 & 2.10 & -0.71 & -3.88 &  -6.86 &  6.18 & 6.11 & \multicolumn{2}{c}{...} &  6.74 & 6.64 & \multicolumn{2}{c}{...} &       &      & \multicolumn{2}{c}{...}  &       &       \\
8194.82 & 2.10 &  0.49 & -4.97 &  -7.23 &  6.76 & 6.23 & \multicolumn{2}{c}{...} &       &      & \multicolumn{2}{c}{...} &  6.52 & 6.23 & \multicolumn{2}{c}{...}  &       &  \\
8194.79 & 2.10 & -0.46 & -4.97 &  -7.23 &       &      &          &              &       &      &      &      &       &      &       &      &      &    \\       
\multicolumn{19}{l}{Sr\ii} \\
4077.71 & 0.0  &  0.15 & -6.22 &  -7.71 &  3.30 & 3.37 & \multicolumn{2}{c}{...} &  4.24 & 4.38 & \multicolumn{2}{c}{...} &  2.89 & 3.48 & \multicolumn{2}{c}{...} &  &   \\        
\multicolumn{19}{l}{Ba\ii} \\
4554.03 & 0.0  &  0.17 & -5.31 &  -7.73 &  2.78 & 2.69 & \multicolumn{2}{c}{...} &  3.58 & 3.68 & \multicolumn{2}{c}{...} &  2.75 & 3.13 & \multicolumn{2}{c}{...} &      &       \\
\noalign{\smallskip}\hline \noalign{\smallskip}
\multicolumn{19}{l}{{\bf Notes.} $\Gamma_4/N_{\rm e}$ and $\Gamma_6/N_{\rm H}$ in rad/s$\cdot$cm$^3$. This table is available in its entirety in a machine-readable form in the online journal.} \\
\multicolumn{19}{l}{A portion is shown here for guidance regarding its form and content. }    \\
\noalign{\smallskip} \hline
	\end{tabular} 
\end{table*}

\begin{table*}
	\centering
	\caption{Mean LTE (L) and NLTE (N) abundances, $\eps{}$, for the sample stars.  }
	\label{tab:abundances}
	\begin{tabular}{lrcccccccc} %
		\hline\hline \noalign{\smallskip}
Atom &      & HD~32115  & HD~73666  & Vega     & HD~72660  & HD~145788 & Sirius   & 21 Peg   & $\pi$ Cet   \\
		\hline \noalign{\smallskip}
He$^1$ & N &        & 10.94(0.03) & 10.99(0.04) & 10.72(0.04) &        & 10.82(0.03) & 10.90(0.02) & 10.94(0.01) \\
\multicolumn{2}{r}{[He/H]} & & 0.01 & 0.06      & -0.21       &        & -0.11       & -0.03       & 0.01      \\
C   & L & 8.55(0.22) & 8.62(0.16) & 8.65(0.28) & 8.00(0.12) & 8.37(0.08) & 7.73(0.14) & 8.36(0.23) & 8.41(0.11) \\
    & N & 8.45(0.08) & 8.57(0.08) & 8.34(0.13) & 8.02(0.08) & 8.32(0.07) & 7.71(0.14) & 8.38(0.09) & 8.45(0.09) \\
\multicolumn{2}{r}{[C/H]}  &  0.06 ( 9) &  0.18 (19) & -0.05 (19) & -0.37 ( 6) & -0.07 ( 5) & -0.69 (14) & -0.01 (25) & 0.06 (9) \\
O  & L &  9.35(0.36) &  9.82(0.41) &  9.65(0.36) &  9.30(0.43) &  8.90(0.06) &  9.17(0.41) &  9.85(0.43) &  9.94(0.44) \\
   & N &  8.86(0.11) &  8.90(0.08) &  8.60(0.05) &  8.43(0.07) &  8.75(0.02) &  8.44(0.05) &  8.63(0.02) &  8.72(0.02) \\
\multicolumn{2}{r}{[O/H]} &  0.13 ( 6) &  0.17 ( 6) & -0.13 ( 6) & -0.30 ( 6) &  0.02 ( 3) & -0.29 ( 5) & -0.10 ( 6) & -0.01 ( 6) \\
Ne  & L &     &      &       &       &      &      &  8.16(0.04) &  8.35(0.07) \\
    & N &     &      &       &       &      &      &  8.00(0.05) &  8.07(0.06) \\      
\multicolumn{2}{r}{[Ne/H]} &            &            &            &            &            &            & -0.05 ()   & 0.02 (15) \\
Na & L &  6.47(0.38) &  6.76(0.20) &             &  6.76(0.02) &             &  6.80(0.15) &  6.45(0.11) &  6.71(0.06) \\
   & N &  6.17(0.07) &  6.47(0.01) &             &  6.66(0.02) &             &  6.54(0.02) &  6.24(0.01) &  6.37(0.05) \\
\multicolumn{2}{r}{[Na/H]} & -0.12 ( 6) &  0.18 ( 4) &            &  0.37 ( 2) &            &  0.25 ( 3) & -0.05 ( 2) &  0.08 ( 3) \\
Mg & L &  7.61(0.08) &  7.87(0.17) &  7.09(0.07) &  7.92(0.16) &  7.78(0.18) &  7.66(0.18) &  7.61(0.15) &  7.62(0.11) \\
   & N &  7.58(0.06) &  7.70(0.04) &  7.05(0.04) &  7.77(0.06) &  7.57(0.07) &  7.51(0.06) &  7.50(0.04) &  7.57(0.05) \\
\multicolumn{2}{r}{[Mg/H]} &  0.04 (12) &  0.16 (22) & -0.49 (12) &  0.23 (21) &  0.03 (10) & -0.03 (17) & -0.04 (21) &  0.03 (16) \\
Si & L &  7.66(0.18) &  7.81(0.25) &  7.08(0.17) &  7.86(0.20) &  7.90(0.19) &  7.84(0.18) &  7.60(0.28) &  7.62(0.24) \\
   & N &  7.59(0.15) &  7.67(0.16) &  6.92(0.15) &  7.83(0.09) &  7.60(0.12) &  7.72(0.11) &  7.50(0.13) &  7.75(0.13) \\
\multicolumn{2}{r}{[Si/H]} &  0.06 (20) &  0.14 (14) & -0.61 ( 8) &  0.30 (20) &  0.07 ( 7) &  0.19 (11) & -0.03 (18) &  0.22 (21) \\
Ca & L &  6.53(0.29) &  6.47(0.13) &  5.82(0.10) &  6.58(0.11) &  6.43(0.10) &  5.88(0.08) &  5.95(0.17) &  5.91(0.59) \\
   & N &  6.39(0.09) &  6.52(0.05) &  6.03(0.10) &  6.62(0.09) &  6.75(0.07) &  6.09(0.05) &  6.30(0.15) &  6.40(0.00) \\
\multicolumn{2}{r}{[Ca/H]} &  0.08 (36) &  0.21 (21) & -0.28 (10) &  0.31 (31) &  0.44 (13) & -0.22 (12) &  -0.01 ( 7) &  0.09 ( 3) \\
Sc & L &  3.22(0.10) &  3.02(0.02) &             &  2.63(0.05) &  3.05(0.04) &  1.99(0.11) &  2.60(0.05) &  2.61(0.10) \\
\multicolumn{2}{r}{[Sc/H]} &  0.15 ( 8) & -0.05 ( 6) &            & -0.43 ( 6) & -0.02 ( 6) & -1.08 ( 4) & -0.47 ( 6) & -0.46 ( 1) \\
Ti & L &  4.73(0.06) &  5.27(0.17) &  4.50(0.02) &  5.49(0.11) &  5.28(0.15) &  5.20(0.06) &  4.80(0.05) &  4.63(0.09) \\
   & N &  4.76(0.05) &  5.19(0.08) &  4.50(0.02) &  5.45(0.08) &  5.23(0.07) &  5.15(0.04) &  4.80(0.04) &  4.90(0.08) \\
\multicolumn{2}{r}{[Ti/H]} & -0.17 (15) &  0.26 ( 8) & -0.43 ( 6) &  0.52 (41) &  0.30 (32) &  0.22 ( 6) & -0.13 (46) & -0.03 (11) \\
Fe$^2$ & L &  7.56(0.11) &  7.70(0.10) &  7.06(0.30) &  8.08(0.16) &  7.79(0.10) &  7.99(0.06) &  7.53(0.07) &  7.39(0.08) \\
   & N &  7.55(0.11) &  7.70(0.10) &  7.05(0.17) &  8.10(0.16) &  7.76(0.07) &  7.98(0.06) &  7.51(0.07) &  7.38(0.10) \\
\multicolumn{2}{r}{[Fe/H]} &  0.09      &  0.24      & -0.41      &  0.67      &  0.30      &  0.52      &  0.05      & -0.08      \\
Sr & L &  3.33(0.04) &  3.00(0.04) &  2.03(0.01) &  4.20(0.06) &  3.37(0.04) &  3.53(0.00) &  2.89(0.00) &  2.64(0.10) \\
   & N &  3.28(0.04) &  3.38(0.00) &  2.71(0.01) &  4.35(0.04) &  3.28(0.01) &  3.83(0.04) &  3.49(0.01) &  2.97(0.10) \\
\multicolumn{2}{r}{[Sr/H]} &  0.38 ( 2) &  0.48 ( 2) & -0.19 ( 2) &  1.45 ( 2) &  0.38 ( 2) &  0.93 ( 2) &  0.59 ( 2) &  0.07 ( 1) \\
Zr & L &  2.75(0.04) &  2.82(0.01) &  2.17(0.10) &  3.92(0.17) &  2.87(0.11) &  3.39(0.13) &  2.41(0.10) &             \\
   & N &  2.82(0.03) &  3.01(0.00) &  2.42(0.10) &  3.85(0.11) &  2.93(0.09) &  3.40(0.13) &  2.92(0.10) &             \\
\multicolumn{2}{r}{[Zr/H]} &  0.25 ( 3) &  0.44 ( 2) & -0.15 ( 1) &  1.28 ( 8) &  0.36 ( 2) &  0.84 ( 2) &  0.35 ( 1) &            \\
Ba & L &  2.64(0.17) &  2.86(0.05) &  1.70(0.04) &  3.54(0.03) &  2.90(0.07) &  3.58(0.06) &  2.78(0.05) &             \\
   & N &  2.47(0.07) &  3.11(0.04) &  2.14(0.02) &  3.67(0.03) &  2.90(0.05) &  3.74(0.06) &  3.16(0.05) &             \\
\multicolumn{2}{r}{[Ba/H]} &  0.29 ( 3) &  0.93 ( 5) & -0.04 ( 3) &  1.49 ( 5) &  0.72 ( 3) &  1.56 ( 5) &  0.98 ( 3) &            \\
Nd & L &  1.31(0.01) &  1.94(0.03) &             &  2.74(0.04) &             &  2.94(0.03) &  1.96(0.00) &             \\
   & N &  1.43(0.01) &  1.94(0.03) &             &  2.73(0.04) &             &  2.93(0.03) &  1.95(0.01) &             \\
\multicolumn{2}{r}{[Nd/H]} & -0.04 ( 2) &  0.47 ( 2) &            &  1.26 ( 2) &            &  1.46 ( 2) &  0.48 ( 2) &            \\
\noalign{\smallskip}\hline \noalign{\smallskip}
\multicolumn{10}{l}{{\bf Notes.} $^1$ from \citet{2018AstL...44..621K}, $^2$ Sitnova (2020, in prep). The numbers in parentheses are } \\
\multicolumn{10}{l}{ the dispersions in the single line measurements around the mean and the numbers of used lines. } \\
\noalign{\smallskip} \hline
	\end{tabular} 
\end{table*}  

\begin{table}
	\centering
	\caption{LTE and NLTE abundances, $\eps{}$, of $\iota$ Her.}
	\label{tab:iHer}
	\begin{tabular}{lccrr} %
		\hline\hline \noalign{\smallskip}
Species & LTE & NLTE & n$_l$ & [X/H] \\ 
\noalign{\smallskip}\hline \noalign{\smallskip}
He\ione $^1$ &      & 10.91(0.02) &    & -0.02 \\ 
C\ii  &  8.58(0.26) &  8.43(0.10) & 14 &  0.04 \\
O\ione & 9.94(0.46) & 8.70(0.02)  &  3 & -0.03 \\
Ne\ione &  8.65(0.14) &  8.04(0.04) & 18 & -0.01 \\
Mg\ii &  7.61(0.14) &  7.55(0.07) &  9 &  0.01 \\
Si\iii &  7.79(0.05) &  7.54(0.07) &  4 &  0.01 \\
Ca\ii &  6.78(0.10) &  6.59(0.10) &  1 &  0.28 \\
Fe\ii $^2$ &  7.26(0.04) &  7.40(0.03) &  6 & -0.06 \\
Fe\iii & 7.46(0.06) &             & 21 &  0.00 \\
\noalign{\smallskip}\hline \noalign{\smallskip}
\multicolumn{5}{l}{{\bf Notes.} $^1$ from \citet{2018AstL...44..621K}, } \\
\multicolumn{5}{l}{ $^2$ Sitnova (2020, in prep). } \\
\noalign{\smallskip} \hline
	\end{tabular}
\end{table}

\section{Abundance results}\label{sect:abund}

The LTE and NLTE abundances from individual lines of O\ione, Na\ione, Sr\ii, Zr\ii -\iii, Ba\ii, and Nd\ii -\iii, as well as the LTE abundances from lines of Sc\ii\ 
are presented in Table~\ref{tab:linelist} for each star. We use the abundance scale where $\eps{H}$ = 12.
Tables~\ref{tab:abundances} and \ref{tab:iHer} present average LTE and NLTE abundances of the chemical elements studied in this paper, as well as of He, C, Ne, Mg, Si, Ca, and Ti from our previous papers \citep[][respectively]{2018AstL...44..621K,2016MNRAS.462.1123A,2020ApJ...896...59A,2018ApJ...866..153A,2020MNRAS.493.6095M,2018MNRAS.477.3343S,sitnova_ti}. 
Abundances of C and Mg in HD~32115 and HD~145788, which are missing in the papers of \citet{2016MNRAS.462.1123A} and \citet{2018ApJ...866..153A}, and abundance of Ti in Vega, which is missing in  \citet{sitnova_ti}, were derived in this study.
Errors of the average abundances were calculated as the dispersions in the single line measurements around the mean: $\sigma_l = \sqrt{\Sigma(\overline{x}-x_i)^2 / (N_l-1)}$. Here, $N_l$ is the total number of used lines. If case of $N_l$ = 1, $\sigma_l = 0.1$~dex was adopted.
In order to compute the [X/H] values, we employ the solar system abundances of \citet{Lodders2009}. 

Detailed NLTE analysis of Fe\ione -\ii\ is in progress (Sitnova 2020, in prep), and we provide here preliminary average abundances. The Fe\iii\ based LTE abundance of $\iota$~Her was calculated using 21 lines from the list of \citet{2012A&A...539A.143N}. Our average LTE abundance is lower than the NLTE abundance of NP12, by 0.14~dex. 

Figures~\ref{fig:normal}, \ref{fig:am}, and \ref{fig:vega} display the LTE and NLTE abundances [X/H] for different chemical species in the individual stars.

\subsection{Oxygen}

We preferred to use a common line list for all the sample stars and selected the two triplets, O\ione\ 6155, 6156, 6158\,\AA\ and O\ione\ 7771, 7774, 7775\,\AA. For HD~145788, only the first triplet was used because the IR lines are not covered by the observed spectrum. 

The NLTE abundance corrections are negative for the O\ione\ lines and grow, in absolute value, toward higher $\Teff$. However, in any given star, a magnitude of $\Delta_{\rm NLTE}$ is rather different for the visual and the IR lines. For example, $\Delta_{\rm NLTE}$(O\ione\ 6155\,\AA) varies between $-0.03$~dex (HD~32115) and $-0.22$~dex ($\iota$~Her), while $\Delta_{\rm NLTE}$(O\ione\ 7771\,\AA) between $-0.68$ and $-1.67$~dex. For each star, NLTE leads to consistent abundances from different lines, with an abundance spread of no more than 0.17~dex (HD~32115), while the difference in LTE abundances reaches 1.3~dex ($\pi$~Cet). 

\subsection{Sodium}

The Na\ione\ 5889, 5895\,\AA\ resonance lines are observed in a broad temperature range, up to $\Teff$ = 12800~K. However, the stellar lines can be blended by the interstellar medium (ISM) and/or telluric lines. This is clearly the case in Vega and HD~145788. In the same stars, we could not measure also Na\ione\ 5682, 5688\,\AA\ due to affecting by the telluric lines.

Abundances derived from the resonance lines appear higher compared with that from the subordinate lines, even in the stars without signatures of the ISM absorption. NLTE reduces the abundance difference (resonance -- subordinate), but does not cancel it completely, leaving 0.09~dex (Sirius) to 0.30~dex (21~Peg). Only for $\pi$~Cet, the NLTE abundances from Na\ione\ 5889, 5895\,\AA\ and Na\ione\ 8194\,\AA\ are consistent, within 0.05~dex. In their NLTE analysis of about 120 A-type main-sequence stars, \citet{2009PASJ...61.1165T} conclude that the Na\ione\ resonance lines cannot be used as a reliable abundance indicator. We support this conclusion and do not account for the resonance lines in the average abundances. The exception is $\pi$~Cet.

In five stars, we could measure the IR lines at 8183 and/or 8194\,\AA. NLTE leads to fairly consistent abundances from different subordinate lines, while the difference in LTE abundances between Na\ione\ 8183, 8194\,\AA\ and Na\ione\ 5682, 5688\,\AA\ amounts to 0.16~dex (21~Peg) to 0.5~dex (HD~32115).

\subsection{Scandium}\label{sect:scandium}

The NLTE calculations for Sc\ii\ were only performed in the literature for late type stars \citep{Zhang2008_sc,lick_paperII} and resulted in slightly positive NLTE abundance corrections. The ionization energy of Sc\ii\ is $\chi_{\rm ion} \simeq$ 12.8~eV. For $\Teff >$ 9000~K, Sc\ii\ is effectively ionized in the atmosphere and is expected to be subject to overionization, similarly to that for Zr\ii, which has close $\chi_{\rm ion} \simeq$ 13.1~eV. The NLTE abundance corrections computed for lines of Zr\ii\ can be used to get an idea of magnitudes of the NLTE abundance corrections for lines of Sc\ii. 

\subsection{Strontium}

Only the Sr\ii\ 4077 and 4215\,\AA\ resonance lines were measured in spectra of our eight stars. In the hottest star, $\iota$~Her, these lines cannot be extracted from noise. In HD~32115, the lines are very strong, with $EW >$ 260~m\AA, and probably cannot be reliable abundance indicators.

\subsection{Zirconium}

For our seven stars with $\Teff \le$ 10400~K, we could use one to three lines of Zr\ii\ in the visible spectral range. 
Six nearly clean lines of Zr\iii\ were found in the UV spectrum of HD~72660 and the only suitable line of Zr\iii, at 1790\,\AA, in Sirius. For HD~72660, NLTE leads to fairly consistent abundances from the two ionization stages, while the difference in LTE abundances amounts to $\eps{}$(Zr\ii\ - Zr\iii) = $-0.28$~dex. Using one line of Zr\ii\ and one line of Zr\iii\ in Sirius, we obtain a discrepancy between the two ionization stages, which, in LTE and NLTE, is very similar in absolute value, but of opposite sign, of $-0.18$ and 0.19~dex, respectively. 

\subsection{Barium}

In each star with $\Teff \le$ 10400~K, we measured the Ba\ii\ 4554 and 4934\,\AA\ resonance lines and one to three lines of the Ba\ii\ $5d-6p$ triplet. In HD~32115, the resonance lines are very strong, with $EW \simeq$ 200~m\AA, and probably cannot be reliable abundance indicators. We obtained an abundance discrepancy between the resonance and subordinate lines of 0.18 and 0.22~dex, in LTE and NLTE, respectively, and excluded Ba\ii\ 4554 and 4934\,\AA\ from calculating the Ba mean abundance of HD~32115.
In contrast, consistent abundances from the resonance and subordinate lines were found in the remaining six stars, with an abundance difference of 0.02 to $-0.12$~dex in different stars. 

\subsection{Neodimium}

We could derive the Nd abundance for five stars of our sample. Two lines of Nd\ii\ were measured in the coolest star, HD~32115, and two lines of Nd\iii\ in each of the four stars, which reveal enhancements of the heavy elements beyond the iron group. 

\begin{figure*}
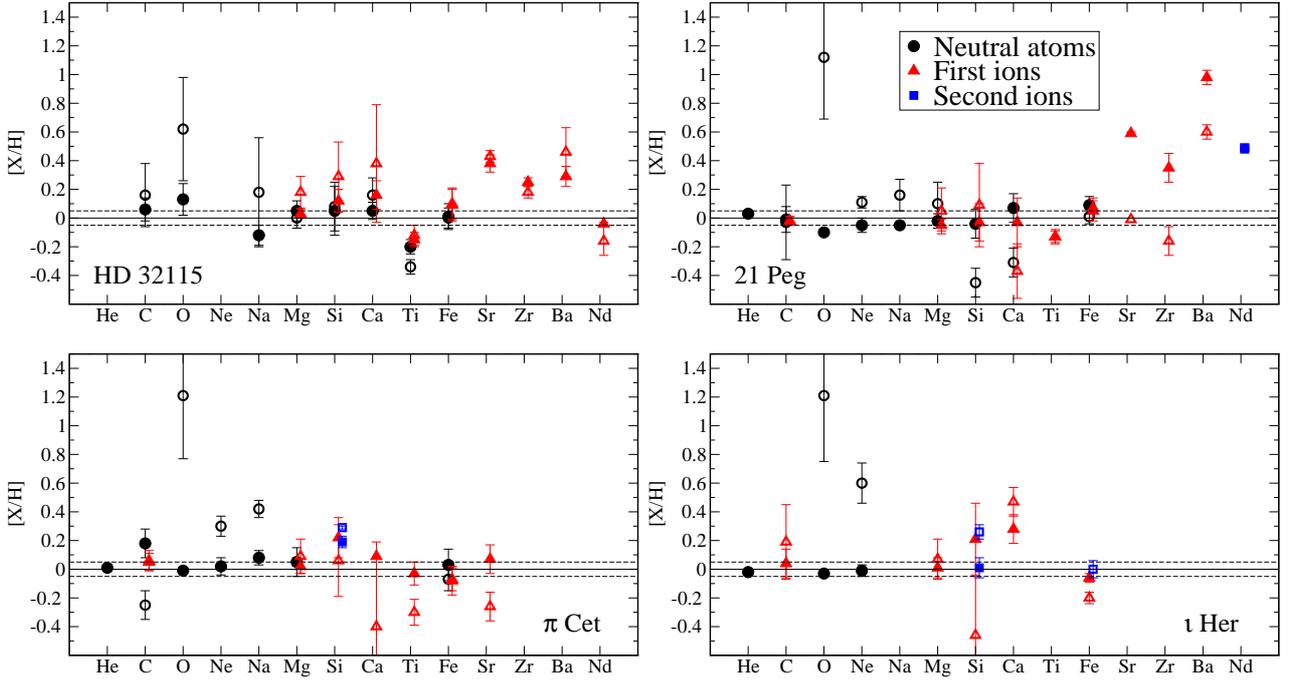

 \begin{minipage}{170mm}
\centering
	\includegraphics[width=0.49\textwidth, clip]{hd32115_abund_f.eps}
	\includegraphics[width=0.49\textwidth, clip]{21peg_abund.eps}

\vspace{3mm}
	\includegraphics[width=0.49\textwidth, clip]{piCet_abund.eps}
	\includegraphics[width=0.49\textwidth, clip]{iHer_abund_f.eps}
    \caption{NLTE (filled symbols) and LTE (open symbols) [X/H] abundances from lines of neutral atoms (circles), first ions (triangles), and second ions (squares) in HD~32115, 21~Peg, $\pi$~Cet, and $\iota$~Her. The short-dashed lines indicate a typical uncertainty of 0.05~dex in the solar system abundances of \citet{Lodders2009}.}
    \label{fig:normal}
\end{minipage}
\end{figure*}

\begin{figure*}
 \begin{minipage}{170mm}
\centering
	\includegraphics[width=0.49\textwidth, clip]{Sirius_abund_f.eps}
	\includegraphics[width=0.49\textwidth, clip]{HD72660_abund.eps}

\vspace{3mm}
	\includegraphics[width=0.49\textwidth, clip]{HD145788_abund.eps}
	\includegraphics[width=0.49\textwidth, clip]{HD73666_abund.eps}
    \caption{The same as in Fig.~\ref{fig:normal} for Sirius, HD~72660, HD~145788, and HD~73666.}
    \label{fig:am}
\end{minipage}
\end{figure*}

\begin{figure}
 \begin{minipage}{85mm}
\centering
	\includegraphics[width=0.99\textwidth, clip]{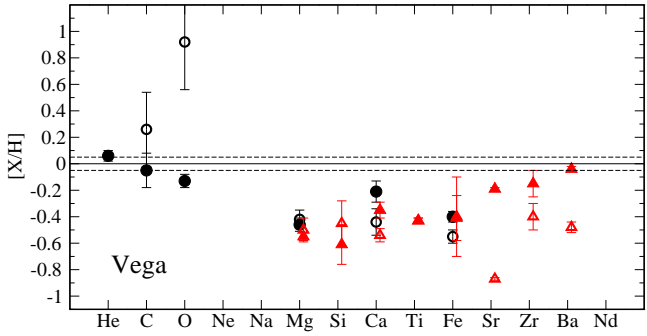}
    \caption{The same as in Fig.~\ref{fig:normal} for Vega.}
    \label{fig:vega}
\end{minipage}
\end{figure}

\section{Discussion}\label{sect:pattern}

In this section, we discuss common features and distinctions in element abundance patterns of our sample stars, which belong to the groups of superficially normal, Am, and $\lambda$~Bootis type stars. We inspect carefully an element abundance pattern of our Blue Straggler star and attempt to understand the nature of HD~145788, with an unidentified type of chemical peculiarity.

\subsection{Superficially normal stars}

For each of our four superficially normal stars, the NLTE abundances of the elements in the range from He to Fe were found to be consistent with the solar values of \citet{Lodders2009}, within 0.1~dex, as seen from Fig.~\ref{fig:normal}. There are following three exceptions.
(i) Titanium is subsolar in HD~32115, independent of NLTE or LTE, with [Ti/H] = $-0.17\pm0.05$ from the NLTE calculations of 15 lines of Ti\ione\ and Ti\ii.
(ii) Silicon is supersolar in $\pi$~Cet, with [Si/H] = $0.22\pm0.13$ in NLTE (21 lines of Si\ii\ and Si\iii). LTE leads to the lower Si enhancement, but substantially larger line-to-line scatter: [Si/H] = $0.09\pm0.24$.
(iii) The only calcium absorption line, Ca\ii\ 3933\,\AA, was measured in $\iota$~Her, and it suggests an enhancement of calcium, with [Ca/H] = 0.28 in NLTE and even larger value in LTE, by 0.19~dex. 

We stress that not only absorption, but also emission lines of C\ione\ (21~Peg, $\pi$~Cet, $\iota$~Her), Si\ii\ ($\iota$~Her), and Ca\ii\ ($\iota$~Her) were well reproduced in our NLTE calculations with classical plane-parallel model atmospheres \citep{2016MNRAS.462.1123A,2020MNRAS.493.6095M,2018MNRAS.477.3343S} providing one more evidence for a status of these stars as normal stars.

The elements beyond Fe are worth to be discussed separately.
Each of the four elements Sr, Zr, Ba, and Nd in 21~Peg is enhanced relative to its solar abundance, by more than 0.3~dex. In HD~32115, Zr and Ba are also supersolar, although to a less extent, with [Zr/H] = 0.25 and [Ba/H] = 0.29~dex, while [Nd/H] is close to the solar value. The Sr\ii\ 4077 and 4215\,\AA\ lines in HD~32115 suggest overabundance of Sr, however, they are rather strong and the obtained [Sr/H] = 0.38 cannot be considered as a reliable value. 
 Only Sr was measured in $\pi$~Cet, and it has a normal (solar) abundance.
No element beyond Fe was detected in the hottest star, $\iota$~Her.

In the following, we investigate whether the obtained enhancement of the neutron-capture (n-capture) elements in 21~Peg can be explained by errors of the abundance determinations, or such a phenomenon is characteristic of A and late B type stars with low rotational velocities. 

\subsubsection{Uncertainties in the NLTE abundances of heavy elements}

\begin{table}
	\centering
	\caption{Changes in the NLTE abundances ($\Delta$, dex) of Sr, Zr, Ba, and Nd in 21~Peg caused by variations in atmospheric parameters and atomic data.}
	\label{tab:uncertainties}
	\begin{tabular}{llrrrr} %
	\hline\hline \noalign{\smallskip}
        &     & Sr     & Zr     & Ba     & Nd     \\ 
\noalign{\smallskip}\hline \noalign{\smallskip}
\multicolumn{6}{l}{ Atmospheric parameters: } \\
 $\Teff$, $-200$~K & $\Delta_{\Teff}$ & $-0.17$ & $-0.19$ & $-0.17$ & $-0.05$ \\ 
 $\logg$, +0.1    & $\Delta_{\logg}$  & $-0.06$ & $-0.07$ & $-0.06$ &  0.03 \\ 
\multicolumn{6}{l}{Photoionizations:} \\
 hydrogenic    & $\Delta_{\rm RBF}$   & $-0.10$ &         & $-0.16$ &       \\ 
\multicolumn{6}{l}{Electron impact ionizations:} \\
 rates $\times$ 10 & $\Delta_{\rm CBF}$ & $-0.05$ & $-0.01$ & $-0.09$ & $< 0.01$  \\ 
\multicolumn{6}{l}{Electron impact excitations:} \\
 vReg, $\Upsilon$ = 1 & $\Delta_{\rm CBB}$ & $-0.01$ &         & 0.01    &     \\
 & $\sigma_{tot}$     &  0.22 &  0.20 &  0.26 &  0.06 \\ 
 & $\sigma_l$ &  0.01 &       &  0.05 &  0.01 \\
\multicolumn{2}{l}{Average $\Delta_{\rm NLTE}$} & 0.60 & 0.51 & 0.38 & $-0.01$ \\
\noalign{\smallskip}\hline 
	\end{tabular}     
\end{table}              

We checked a sensitivity of the NLTE abundances from lines of Sr\ii, Zr\ii, Ba\ii, and Nd\iii\ in 21~Peg to variations in atmospheric parameters and atomic data used in the NLTE calculations. Table~\ref{tab:uncertainties} summarises results of our tests. 

\citet{2009AA...503..945F} estimate the uncertainties in $\Teff$, $\logg$, and $\xi_t$ of 21~Peg as 200~K, 0.1~dex, and 0.5~\kms, respectively. Since enhancements of Sr and Zr are due to large positive NLTE abundance corrections, we have chosen to decrease $\Teff$, but increase $\logg$ in our test calculations. We do not indicate abundance shifts due to variations in $\xi_t$ because all the lines under investigation are weak, with $EW < 30$~mA.

We assumed that a maximal influence of variations in photoionization cross sections on the NLTE results can be estimated when replacing accurate data from the DBSR method (see Sect.~\ref{sect:sr2}) with the hydrogenic cross sections. Such estimates can be done for Sr\ii\ and Ba\ii. 
Since the NLTE effects were reduced, when using hydrogenic photoionization cross sections, we assume that our calculations for Zr\ii\ provided the lower limit for the Zr NLTE abundance. Variations in photoionization cross sections cannot affect the NLTE abundances derived from lines of Nd\iii\ because Nd\iii\ is a majority species in the atmosphere of 21~Peg.

For Sr\ii, Zr\ii, and Ba\ii, the main NLTE mechanism is the overionization. The NLTE effects can be reduced, by increasing the electron impact ionization rates. We chosen a scaling factor of 10.

For Sr\ii\ and Ba\ii, we also checked the changes in the derived NLTE abundances, when replacing  accurate electron-impact excitation data from \citet{2002MNRAS.331..875B} for Sr\ii\ and from \citet{1974PhRvA..10..141C} and \citet{1981eabs.book.....S} for Ba\ii\ with the formula of \citet{Reg1962} for all allowed transitions and $\Upsilon$ = 1 for all forbidden transitions. The corresponding string in Table~\ref{tab:uncertainties} is denoted as vReg, $\Upsilon$ = 1.

It can be seen from Table~\ref{tab:uncertainties} that, for Sr\ii\ and Ba\ii, a variation in photoionization cross sections has the bigger effect on the departures from LTE compared with that for a variation in collisional data.
The total impact of varying each of the four parameters, $\sigma_{tot}$, was computed as the quadratic sum of $\Delta_{\Teff}$, $\Delta_{\logg}$, $\Delta_{\rm RBF}$, and $\Delta_{\rm CBF}$. For comparison, we indicate the abundance error which is characteristic of the line-to-line scatter, $\sigma_l$, and the average $\Delta_{\rm NLTE}$. For none of the four chemical elements, their obtained enhancements cannot be explained by the uncertainties in abundance determinations. Barium and neodimium have substantially supersolar abundances, independent of whether LTE or NLTE abundances are determined. 

\begin{figure}
 \begin{minipage}{85mm}
\centering
	\includegraphics[width=0.99\textwidth, clip]{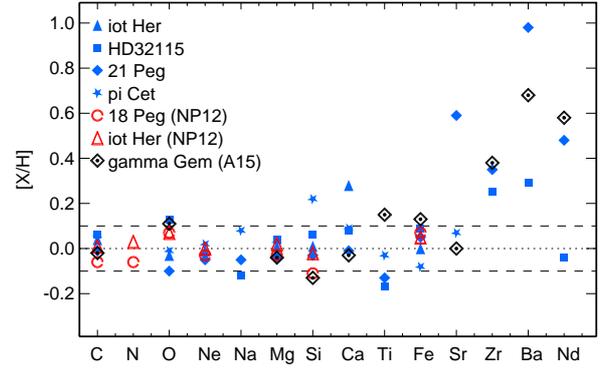}
    \caption{Element abundance patterns of the superficially normal stars. Our NLTE abundances for $\iota$~Her (filled triangles), HD~32115 (squares), 21~Peg (filled rhombi), and $\pi$~Cet (5-pointed stars) compared with NLTE abundances of 18~Peg (circles) and $\iota$~Her (open triangles) from \citet[][NP12]{2012A&A...539A.143N} and LTE abundances of $\gamma$~Gem (rhombi with a small circle inside) from \citet[][A15]{2015PASP..127..340A}. The dashed lines indicate typical abundance error of 0.1~dex. Note that the hotter star the higher abundances of Zr, Ba, and Nd are.}
    \label{fig:normal2}
\end{minipage}
\end{figure}

\subsubsection{Comparison with the literature}

For each of the four stars, a vast number of abundance determinations was made in the literature, mostly under the LTE assumption; see, for example, \citet{1998MNRAS.296..856A} for $\iota$~Her and \citet[][and references therein]{2009AA...503..945F} for 21~Peg and $\pi$~Cet. The NLTE abundances were derived for $\iota$~Her \citep[][C, N, O, Ne, Mg, Si, Fe]{2012A&A...539A.143N} and a limited number of chemical elements (Na, K, Ca, Sr, Ba) in HD~32115 \citep{2002A&A...389..537B}. The latter paper is based on the NLTE methods, which are very similar to that applied in this study. Small differences in the NLTE abundances between the two studies are due to using different observed spectra of HD~32115. For common elements in $\iota$~Her, our NLTE results agree very well with that of NP12.

Element abundance pattern of $\pi$~Cet, as obtained by \citet[][F09]{2009AA...503..945F} under the LTE assumption, observed emission features at the position of the C\ione\ lines in the near infrared and close to the core of H$_\alpha$, and the variations in the line profiles within one day make doubt in a status of $\pi$~Cet as a normal star. The NLTE calculations remove most of the problems of the LTE analysis and provide a firm evidence for $\pi$~Cet being an abundance standard of a normal B-type star. Indeed, \citet{2016MNRAS.462.1123A} showed that the emission in C\ione\ 8335 and 9405\,\AA\ appears due to the departures from LTE and these emission lines form in the atmosphere of $\pi$~Cet. NLTE removes an overabundance of O, Ne, Na, and Mg and an underabundance of Ti obtained by F09.

Compared with detailed LTE abundances reported by F09 for 21~Peg, our NLTE calculations remove an overabundance of Ne and an underabundance of Ca relative to their solar abundances. Based on our NLTE calculations for Zr\ii, we expect positive NLTE abundance corrections for lines of Sc\ii, at the level of 0.5~dex (see Sect.~\ref{sect:scandium}). Solar abundances of Ca and Sc in 21~Peg are in particular important because subsolar Ca as well Sc abundances serve as formal classification criteria of Am stars. We, thus, exclude finally an assumption that 21~Peg could be the hottest known Am star. We also note that NLTE removes a big abundance discrepancy between Si\ii\ and Si\ione\ obtained by F09.

Inspecting the literature leads us to conclude that enhancement of the n-capture elements, as found for 21~Peg and HD~32115, is a widely spread phenomenon among chemically normal A-type stars \citep[][and references therein]{1990A&A...240..331L,1993AA...276..142H,1998MNRAS.296..856A}. 
 In Fig.~\ref{fig:normal2}, we compare our NLTE abundance patterns with the literature data for superficially normal stars. These are $\iota$~Her and 18~Peg from \citet[][NLTE]{2012A&A...539A.143N} and $\gamma$~Gem from \citet[][LTE]{2015PASP..127..340A}. The stars $\iota$~Her and 18~Peg ($\Teff$ = 15800~K) are too hot for reliable measurements of any of the n-capture elements in the visible spectra. The star $\gamma$~Gem, with $\Teff$ = 9150~K, is referred to as a classical superficially normal A-type star. 
 In order to display element abundance pattern of $\gamma$~Gem, we selected the abundances derived by \citet{2015PASP..127..340A} from lines of C\ii, O\ione, Mg\ii, Si\ii, Ca\ione -\ii, Ti\ione, and Fe\ii, for which the NLTE abundance corrections are minor, according to our estimates. For Sr, Zr, and Ba, their NLTE abundances would be higher than the LTE ones, by 0.2 to 0.3~dex. It is evident that 21~Peg, $\gamma$~Gem, and HD~32115 reveal an enhancement of the n-capture elements, and its magnitude correlates with $\Teff$: the hotter star the higher abundances of the heavy elements are.

\begin{figure}
 \begin{minipage}{85mm}
\centering
	\includegraphics[width=0.99\textwidth, clip]{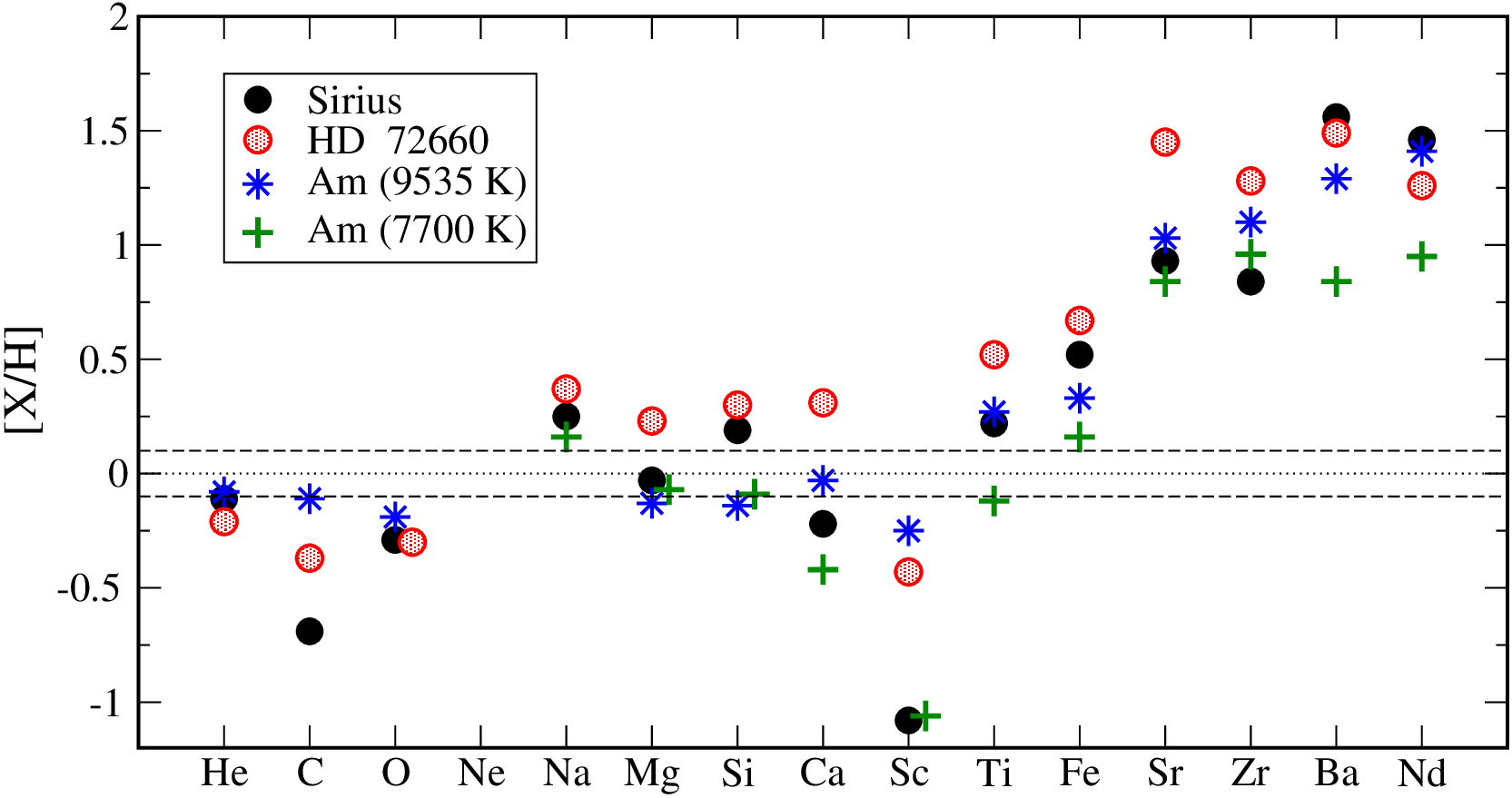}
    \caption{Element abundance patterns of the Am stars. Our sample stars Sirius and HD~72660 compared with $o$~Peg \citep[][$\Teff$ = 9535~K]{2015PASP..127...58A} and 32~Aqr \citep[][$\Teff$ = 7700~K]{1997MNRAS.288..470A}.}
    \label{fig:am1}
\end{minipage}
\end{figure}

\subsection{Am stars}

Sirius and HD~72660 were previously classified as hot Am stars by \citet[][and references therein]{2011A&A...528A.132L} and \citet{1999A&A...341..233V}, respectively. They have similar $\Teff$ and $\logg$, however, HD~72660 has higher abundances of metals in the range from Na to Fe, by 0.1~dex (Si) to 0.5~dex (Ca). 
The NLTE abundances of Sirius and HD~72660 are compared in Fig.~\ref{fig:am1} with the LTE abundances of classical hot Am star $o$~Peg \citep{2015PASP..127...58A} and cool Am star 32~Aqr \citep{1997MNRAS.288..470A}. Both Am stars of our sample 
follow a general abundance trend of the Am stars. They have a deficit of Sc (the key element in Am classification) relative to nearby chemical elements, although of different magnitude for Sirius and HD~72660. This does not make doubt in their status of Am stars, because $o$~Peg and 32~Aqr also reveal rather different underabundances of Sc.

\subsection{HD~73666}

\begin{figure*}
 \begin{minipage}{170mm}
\centering
	\includegraphics[width=0.49\textwidth, clip]{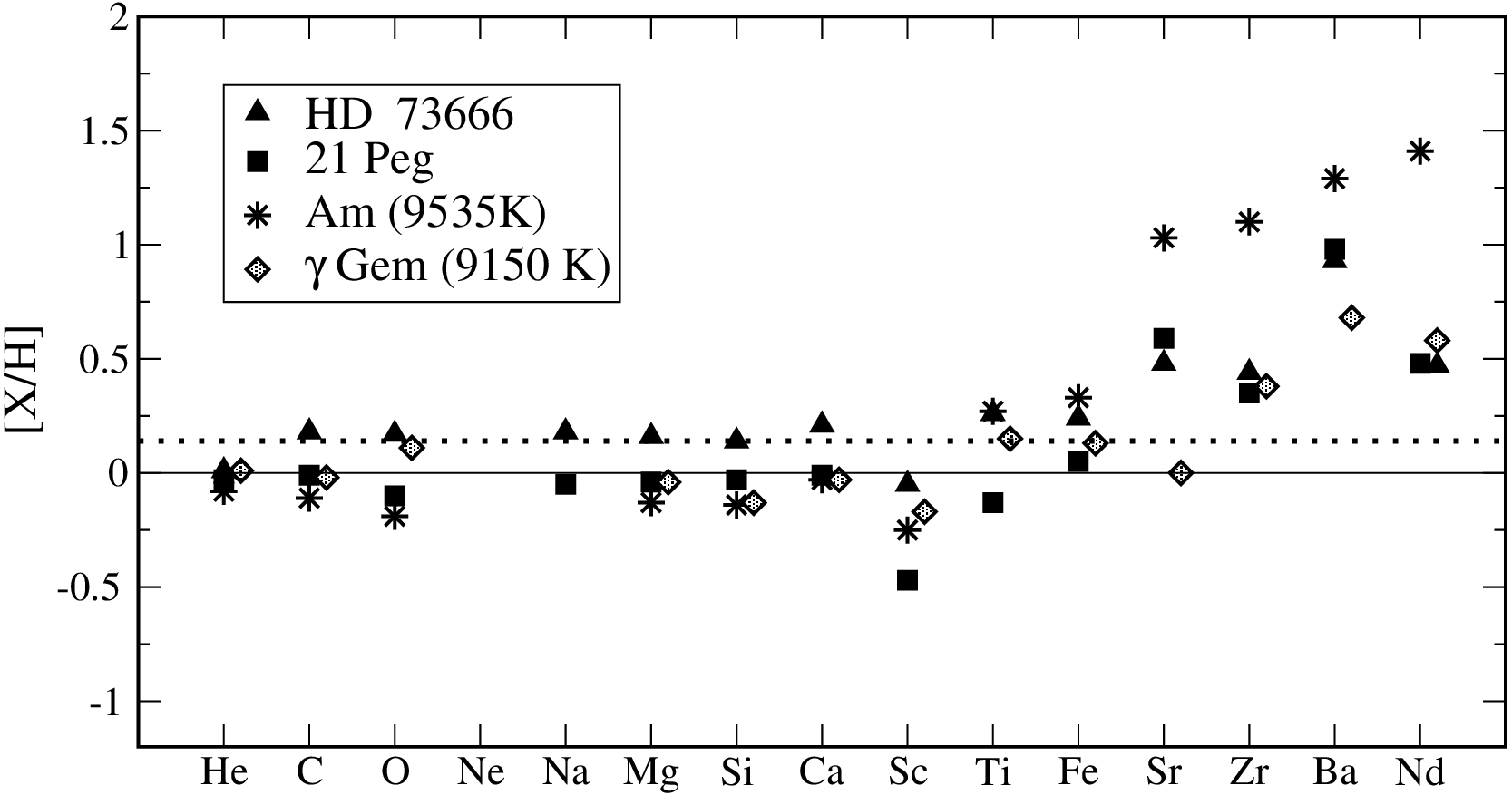}
	\includegraphics[width=0.49\textwidth, clip]{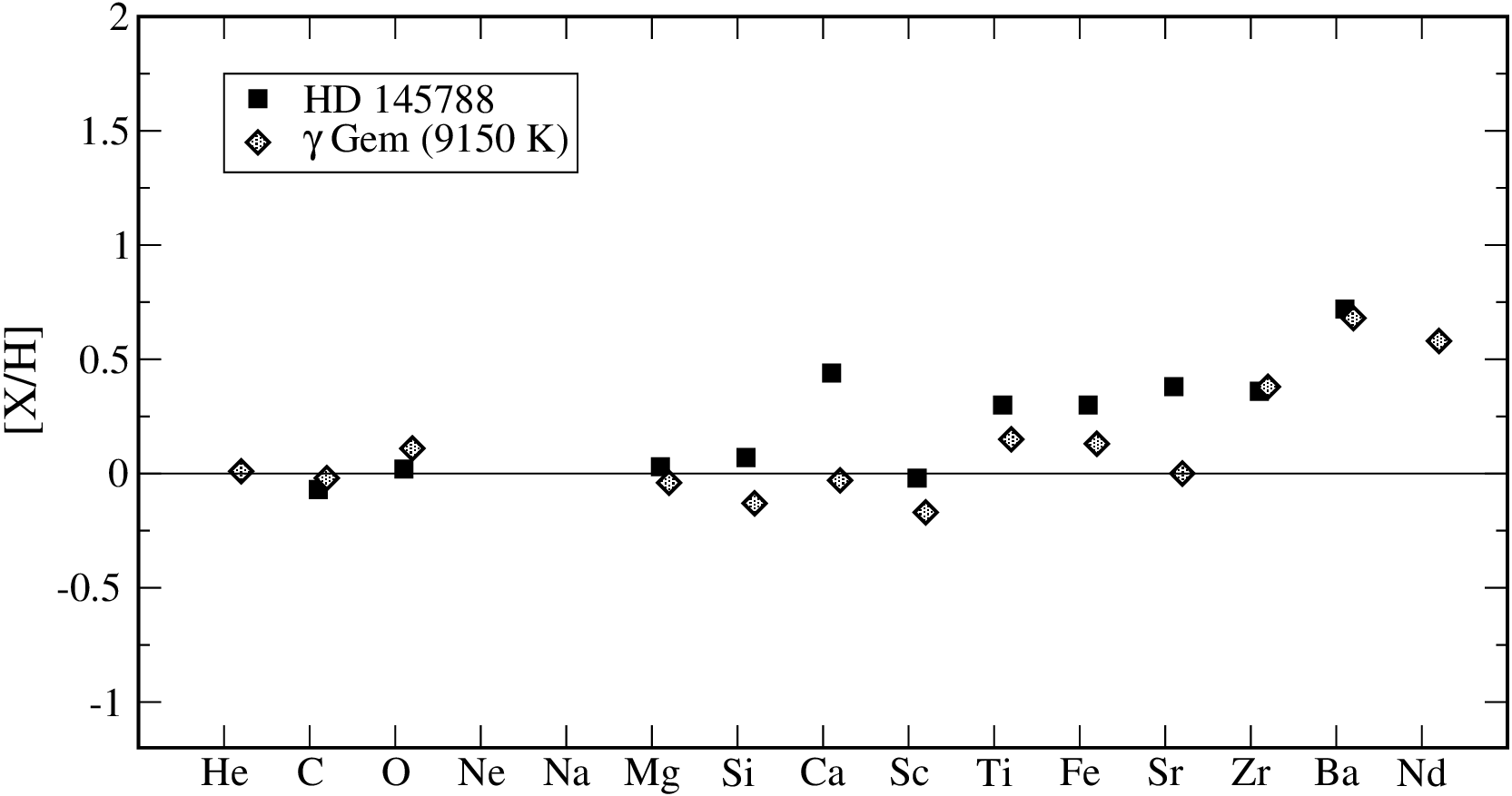}
    \caption{Element abundance patterns of HD~73666 (left panel) and HD~145788 (right panel) compared with that of an Am star $o$~Peg \citep[][$\Teff$ = 9535~K]{2015PASP..127...58A} and superficially normal stars 21~Peg (this work) and $\gamma$~Gem \citep{2015PASP..127..340A}. 
The dotted line indicates a metallicity of the Praesepe open cluster, [Fe/H] = 0.14 \citep{2003AJ....125.1397C}. See text for a discussion of expected NLTE abundance corrections for Sc. }
    \label{fig:hd73666}
\end{minipage}
\end{figure*}

Based on their LTE abundance analysis, \citet{2007AA...476..911F}, refer to HD~73666 as normal A-type star. We now look at this star from the NLTE pespective.
Element abundance pattern of HD~73666 is displayed in Fig.~\ref{fig:hd73666} together with abundances of a classical hot Am star $o$~Peg \citep[][LTE, $\Teff$ = 9535~K]{2015PASP..127...58A} and two superficially normal stars: 21~Peg (this study, NLTE) and $\gamma$~Gem \citep[][LTE, $\Teff$ = 9150~K]{2015PASP..127..340A}. 
Elements from C to Ca in HD~73666 reveal very similar overabundances of 0.14-0.20~dex relative to their solar abundances, in line with the cluster metallicity: [Fe/H] = 0.14 \citep{2003AJ....125.1397C} and 0.11 \citep{2008A&A...483..891F}. There is a hint of the higher enhancement for the heavier elements Ti and Fe, with [Ti/H] = 0.26 and [Fe/H] = 0.24.
 The LTE abundance of Sc reveals a deficit of [Sc/H] = $-0.05$. As discussed in Sect.~\ref{sect:scandium}, we expect the NLTE abundance corrections for lines of Sc\ii\ to be similar to that for Zr\ii. For HD~73666, the difference between NLTE and LTE abundances of Zr amounts to 0.19~dex, and we expect that NLTE would place Sc close to the other elements from C to Fe in the abundance pattern. We also expect that NLTE would largely remove deficit of Sc in 21~Peg, where $\Delta_{\rm NLTE}$ = 0.51~dex for Zr\ii\ 4149~\AA. The comparison stars $\gamma$~Gem and $o$~Peg, most probably, do not have a deficit of Sc as well.  
Abundances of the heavy elements in HD~73666 agree with those for the superficially normal stars 21~Peg and $\gamma$~Gem. 

Compared with the LTE abundance analysis of \citet{2007AA...476..911F}, NLTE reduces overabundances of C, O, and Na to the level characteristic of the other elements and removes substantial abundance discrepancies between Mg\ione\ and Mg\ii\ and between Ca\ione\ and Ca\ii\ increasing, thus, a credit of confidence in the obtained element abundance pattern. Our NLTE abundance analysis provides a firm evidence for HD~73666 being a chemically normal star despite catastrophic processes in the Praesepe open cluster that led to the formation of a Blue Straggler.

 
\subsection{HD~145788}

The star HD~145788 was studied by \citet{2009AA...503..945F} who find an overall enhancement of metals, but do not find the Sc deficit that is a typical classification characteristic of an Am star. They come to the conclusion that, similarly to HD~73666, HD~145788 is a normal star reflecting a chemical composition of the cloud where it formed. Figure~\ref{fig:hd73666} shows the NLTE abundance pattern of this star in comparison with the LTE abundances of $\gamma$~Gem \citep{2015PASP..127..340A}. HD~145788 has close-to-solar abundances of C, O, Mg, and Si, while Ti and Fe are enhanced at the level of 0.3~dex. The higher overabundance was obtained for Ca, [Ca/H] = 0.44. Scandium can be slightly enhanced, if the NLTE effects for Sc\ii\ are similar to that for Zr\ii. Overabundances of the elements heavier than Fe are similar to that for the superficially normal star of similar temperature, $\gamma$~Gem, but substantially lower compared with that for the Am stars of similar temperature, HD~72660 and Sirius.

 The calculated overabundances of Ti to Ba cannot be caused by the NLTE treatments because LTE leads to even higher [X/H] for Ti, Fe, and Sr and the same [Ba/H] (Table~\ref{tab:abundances}). They cannot be due to 
underestimated microturbulent velocity.  We used $\xi_t$ = 1.3~\kms, as derived by \citet{2009AA...503..945F} and proved by \citet{sitnova_ti} in NLTE analysis of the Ti\ii\ lines. 
With our detailed NLTE analysis, we cannot clarify the status of HD~145788.

\subsection{Superficially normal versus Am stars}\label{sect:Discussion}

Careful NLTE abundance study of eight slowly rotating A and B-type stars shows that they have a common feature. This is overabundances of the n-capture elements Sr, Zr, Ba, and Nd, although they are observed at different level in superficially normal and Am stars.
The question arises: how the observed overabundances of heavy elements in A stars are connected with stellar characteristics, such as rotation, magnetic field, binarity, effective temperature?  

\citet{2014A&A...562A..84R} show that the LTE abundances of Y, Zr, and Ba in 21 normal A1-A2 ($\Teff >$ 8900~K) stars are supersolar, with the median values [Y/H] $\simeq$ 0.15, [Zr/H] $\simeq$ 0.5, and [Ba/H] $\simeq$ 0.8. In this temperature range, our NLTE calculations predict positive NLTE abundance corrections for lines of Zr\ii\ and Ba\ii. This means an even greater discrepancy with the solar abundances. Strontium is probably also supersolar, because \citet{2014A&A...562A..84R} obtain [Sr/H] $\simeq -0.25$ in LTE, while the NLTE abundance corrections can be at the level of +0.5~dex for lines of Sr\ii. It is interesting that the comparison sample composed of nine normal A-type stars from the Pleiades and the Ursa Major moving group has the LTE abundances close, on average, to solar abundances of Y-Ba. \citet{2014A&A...562A..84R} note that different rotation velocities can be a source of discrepancies between the two stellar samples because the main sample stars have $v \sin i \le$ 65~\kms, while seven of nine comparison stars have 100 $\le v \sin i \le 200$~\kms.

\citet{2008A&A...483..891F} search for abundance trends with respect to $\Teff$ and $v \sin i$  for three samples of the Am, A- and F-type stars in the Praesepe cluster. All their stars are cooler than 8200~K. This can explain the lower enhancement of Ba, at the level of 0.5~dex, in the normal A-type stars compared with the corresponding value reported by \citet{2014A&A...562A..84R}. The normal A and Am stars have different $v \sin i$, with A-type stars being faster rotators ($v \sin i$ = 75 to 190~\kms) compared with the Am stars ($v \sin i \le$ 70~\kms). The Am stars reveal high abundances of heavy elements, with [Y/H] and [Ba/H] of up to 1.2 and 2~dex, respectively, which decrease with $v \sin i$ and approach the corresponding abundances of the slowest rotators among the normal A-type stars. No clear correlation is found between abundance and $\Teff$ and between abundance and $v \sin i$ for the normal A-type stars. The former can be due to missing the $\Teff >$ 8500~K stars.

\citet{1999A&A...351..247V} make similar analysis for the Hyades open cluster. Their sample of the normal A and Am stars covers similar $\Teff$ range, with the only star of $\Teff >$ 8300~K,  and similar $v \sin i$ range, with the only star of $v \sin i >$ 200~\kms. Figure~8 in \citet{1999A&A...351..247V} shows a hint of correlation between Ba abundance and $v \sin i$, although, in each of three $v \sin i$ ranges, that is, $v \sin i \le$ 50~\kms, 50 $< v \sin i \le 100$~\kms, and $v \sin i >$ 100~\kms, the statistics of stars is poor and a scatter of [Ba/H] exceeds 0.5~dex.


The Am star Sirius possesses a ultra-weak global magnetic field ($<1$~G) \citep{2011A&A...532L..13P}, while no information about magnetic field of HD~72660 is known.
Recently \citet{2020MNRAS.492.5794B} measured a 10~G averaged magnetic field in a superficially normal A star $\gamma$~Gem. Based on some common characteristics of $\gamma$~Gem and Sirius (presence of the magnetic field and belonging to binary systems) and despite their rather different atmospheric chemistry, the authors classify $\gamma$~Gem as definite/potential Am star.
Global magnetic field, together with the SrNd overabundances, are characteristics of Ap stars, but
it is unlikely that the same mechanism produces overabundances of heavy elements in Am and Ap stars. In contrast to the Am stars, where all the elements beyond Sr are overabundant by approximately the same amount, the Ap stars reveal enhancements of Sr and REE's at the level of 2-3~dex, while normal or moderately overabundant Zr and Ba \citep[see Fig.2 in][]{2019MNRAS.488.2343R}.  

We note that no large-scale magnetic field at the level of a few gauss was detected in 21~Peg \citep{2013A&A...554A..61K} and HD~73666 \citep{2007AA...476..911F}, which reveal substantial enhancements of heavy elements. 

Peculiar abundances of the n-capture elements in the atmospheres of superficially normal slowly rotating stars are produced, probably, by the same mechanism as that responsible for the Am phenomenon. However, it operates less efficiently resulting in only moderate enhancements of the n-capture elements and close-to-solar abundances of Fe and the lighter elements.
The question is whether the Am phenomenon is a manifestation of a certain stage of the star's evolution. Can superficially normal A and late B stars with low rotational velocities be nascent Am stars? Or on the contrary, did they already pass the Am stage? Or is there no relationship between normal A and Am stars?  

 We note that, in HD~32115 and 21~Peg, barium has greater overabundance relative to the solar value compared with that for the other heavy elements. The published abundances of heavy elements in young ($<$~2~Gyr) open clusters and associations \citep{2009ApJ...693L..31D,2013AJ....145..107J,2015MNRAS.446.3651M,2015AJ....149...15O,2015MNRAS.454.1976R,2017A&A...598A..86D} find a statistically significant trend of increasing cluster [Ba/Fe] as a function of decreasing cluster age, while all the other n-capture elements, such as Y, Zr, La, Ce, exhibit a solar scaled pattern. Various ideas were suggested to explain extra production of Ba, for example, enhanced s-process production in low-mass (1-1.5~$M_\odot$) AGB stars with highly effective mixing \citep{2009ApJ...693L..31D} or the intermediate neutron-capture process \citep{2015MNRAS.446.3651M}, however, the origin of extra amount of Ba in young stellar systems remains the puzzle. Could it be that the solar value does not represent the modern galactic abundance of Ba and our sample stars, which all are less than 1~Gyr old, formed from the matter enriched with Ba more than the Sun? 

\subsection{Vega}

\citet{1990ApJ...363..234V} identified Vega as a mild $\lambda$~Bootis star. As shown first by \citet{1988A&A...207..112B}, this type stars have essentially normal (solar) abundances of the light elements (C, N, and O) and the underabundances of the heavier elements (Mg, Al, Si, S, Mn, Fe, and Ni). Abundance anomalies are associated with the accretion of metal-poor material from a circumstellar disc or cloud \citep{1990ApJ...363..234V}.

Compared with published NLTE abundances of Vega, this study supplemented its NLTE abundance pattern by Si, Ca, Ti, and Zr.
We obtained only slightly subsolar abundances of C ([C/H] = $-0.05$) and O ([O/H] = $-0.13$) and deficit of Mg, Si, Ca, Ti, and Fe, with [X/H] of $-0.3$ down to $-0.6$~dex. These results are in line with a status of Vega as a mild $\lambda$~Bootis star. Heavy elements Sr, Zr, and Ba turn out to be less underabundant, with [X/H] between $-0.19$ and $-0.04$. 

For common chemical elements (C, O, Mg, Fe, Sr, and Ba), our NLTE abundances agree within 0.01 to 0.10~dex with the earlier determinations of \citet{1986A&A...165..170G,Gigas88,1991A&A...246..644S,1992PASJ...44..649T,1999ARep...43..819B} and \citet{2000A&A...359.1085P,Przybilla_mg,2001A&A...379..936P}. 


\section{Conclusions}\label{sect:Conclusions}

This study presents a comparative analysis of the element abundance patterns of the stellar sample that includes slowly rotating superficially normal, Am, $\lambda$~Bootis type stars and the two stars with unidentified type of chemical peculiarity. The abundance patterns include 14 chemical elements in the range from He to Nd.
In order to increase a confidence in our results and conclusions, we use the stars with well determined atmospheric parameters and high-quality observed spectra available and derive chemical abundances based on the NLTE line formation. The NLTE methods for He\ione, C\ione, O\ione, Ne\ione, Mg\ione -\ii, Si\ione -\ii -\iii, Ca\ione -\ii, and Ti\ione -\ii\ were treated in our earlier studies. 
Here, we present first determinations of the NLTE abundances for Na, Sr, Zr, Ba, and Nd in the sample stars. The exception is Vega, for which the NLTE abundances of Sr and Ba were derived earlier by \citet{1999ARep...43..819B} and \citet{Gigas88}, respectively.

We constructed new model atom of Zr\ii -\iii\ and showed that, in the stellar parameter range with which we concern, NLTE leads to weakened lines of Zr\ii, with positive NLTE abundance corrections of 0.05 up 0.5~dex depending on $\Teff$, while strengthened lines of Zr\iii, with negative NLTE abundance corrections at the level of $-0.1$ to $-0.2$~dex. For HD~72660, we obtained consistent NLTE abundances from lines of the two ionization stages, Zr\ii\ and Zr\iii, while the difference in LTE abundances amounts to $-0.28$~dex. 

The NLTE calculations for Sr\ii\ and Ba\ii\ were performed using for the first time accurate photoionization cross sections, which were calculated for this study with the Dirac $B$-spline $R$-matrix (DBSR) method, in the same approximations as \citet{2010PhRvA..81d3423Z} adopted. In the hot ($\Teff >$ 9000~K) atmospheres, both Sr\ii\ and Ba\ii\ are subject to strong overionization resulting in weakened lines and positive NLTE abundance corrections. Depending on $\Teff$, they amount to 0.35 to 0.59~dex for Sr\ii\ 4077, 4215\,\AA\ and 0.26 to 0.38~dex for Ba\ii\ 4554, 4934\,\AA. 

We obtained that the NLTE abundances of He to Fe in HD~32115 (He was not measured), 21~Peg, $\pi$~Cet, and $\iota$~Her are consistent with the solar abundances of \citet{Lodders2009}, mostly within 0.1~dex, providing a firm evidence for a status of these stars as superficially normal stars. We note that, for $\iota$~Her, our results support the earlier conclusion of \citet{2012A&A...539A.143N}. 
Elements beyond the Fe group 
reveal pronounced enhancements in 21~Peg, at the level of 0.6, 0.35, and 0.5~dex for Sr, Zr, and Nd, respectively. Even larger deviation from the solar abundance, of one order of magnitude, was obtained for Ba. Strontium, zirconium, and barium are also enhanced in HD~32115, but to a less extent. 
 Combining our results with the literature data on an enhancement of the n-capture elements in A-type stars, we conclude that its magnitude correlates with $\Teff$: the hotter star the higher abundances of the heavy elements are.
Only Sr of the n-capture elements was measured for $\pi$~Cet, with the NLTE abundance close to the solar one.
This star can be referred to as an abundance standard of a normal late B-type star.  

In HD~73666 (a member of the Praesepe open cluster, a Blue Straggler), elements from C to Fe reveal an enhancement, which is close to the cluster metallicity: [Fe/H] = 0.14~dex \citep{2003AJ....125.1397C} and 0.11 \citep{2008A&A...483..891F}. Heavy elements Sr, Zr, Ba, and Nd are overabundant, at the level similar to that for the superficially normal A-type stars of close temperature. Our NLTE abundance analysis provides a firm evidence for HD~73666 being a chemically normal star despite catastrophic processes that led to the formation of a Blue Straggler.

Although both Am stars of our sample, Sirius and HD~72660, have similar atmospheric parameters, they differ in chemical abundances. Generally, HD~72660 is more metal rich than Sirius except Ba and Nd which are slightly more abundant in Sirius. The enhancement of Sr, Zr, Ba, Nd reaches 1.5~dex.  

With our detailed NLTE analysis, we cannot clarify a status of HD~145788. This star has close-to-solar abundances of C, O, Mg, and Si, while Ca, Ti, and Fe are enhanced, at the level of 0.4 to 0.3~dex. In the range of heavy elements, beyond Sr, abundance pattern of HD~145788 resembles that for superficially normal A-type stars.  


 We propose that the same mechanism produces enhancements of the n-capture elements in the atmospheres of Am and superficially normal stars with low rotational velocities. However, in the latter type of objects, it operates less efficiently, not affecting  abundances of Fe and the lighter elements and resulting in only moderate enhancements of heavy elements. To understand this mechanism(s) is a challenge for the physics of stars. Observations show that the level of enhancement of heavy elements is higher for the hotter stars and, probably, does not depend on the presence or absence of magnetic field.

Compared with the literature data, the NLTE abundance pattern of Vega was supplemented by Si, Ca, Ti, and Zr. In line with a status of Vega as a mild $\lambda$~Bootis star, C and O have close-to-solar abundances and Mg to Fe are underabundant, at the level of 0.4~dex. Less underabundances, of $-0.19$ to $-0.04$~dex, were found for the heavy elements Sr, Zr, and Ba. 

In every our star with Ba abundance available, [Ba/H] was found to be larger than [X/H] for any other element X. Taking also into account published Ba abundances of young open clusters, we propose that the solar Ba abundance is not representative of the galactic Ba abundance at modern epoch.

{\it Acknowledgments.} 
The authors thank V.~Khalack, A.J.~Korn, and J.~Landstreet for providing observed spectra of HD~72660 and Vega. L.M., T.R., and T.S. thank the Ministry of Science and Higher Education of Russian Federation (project 13.1902.21.0039) for a support of this study.
 This study made use of the ESO UVESPOP, NIST, TOPbase, VALD, ADS\footnote{http://adsabs.harvard.edu/abstract\_service.html}, and R.~Kurucz's databases. The authors thank the referee Luca Fossati for his constructive suggestions and remarks.
 
\section{Data availability}

The data underlying this article will be shared on reasonable request to the corresponding author.

\bibliography{atomic_data,mashonkina,nlte,ab2020,si2018}
\bibliographystyle{mnras}

\label{lastpage}
\end{document}